\documentclass[12pt]{article}

\usepackage{indentfirst}
\usepackage{amsmath}
\usepackage{amssymb}
\usepackage{amsfonts}
\usepackage{amscd}
\usepackage{amsbsy}
\usepackage{amsthm}
\usepackage{latexsym}
\usepackage{verbatim}
\usepackage{graphicx,color} 
\usepackage[dvipsnames]{xcolor}
\usepackage[colorlinks]{hyperref}
\hypersetup{linkcolor=blue,citecolor=blue,urlcolor=blue}
\def\nn{\nonumber}       
\def\beq{\begin{eqnarray}}
\def\eeq{\end{eqnarray}}
\def\ln{\,\mbox{ln}\,}

\def\al{\alpha}
\def\be{\beta}

\def\ga{\gamma}

\def\pa{\partial}

\def\si{\sigma}

\def\th{\theta}

\def\Ga{\Gamma}

\def\La{\Lambda}

\usepackage{float} 
\usepackage{cite}  
\usepackage{titlesec}

\usepackage{ulem} 

\titleformat*{\section}{\large\bfseries}
\titleformat*{\subsection}{\normalsize\bfseries}

\begin{document}

\global\long\def\imsize{0.83\columnwidth}%

\global\long\def\halfsize{0.45\columnwidth}%

\global\long\def\thirdsize{0.3\columnwidth}%

\global\long\def\peq{0.65\columnwidth}%

\begin{center}

{\large\bf
Six-dimensional cosmological models with conformal extensions}
\vskip 4mm

Daniel  M\"{u}ller$\,^{a}$\footnote{
E-mail address:  dmuller@unb.br},
\quad
Sergey G. Rubin$\,^{b}$\footnote{
E-mail address:   sergeirubin@list.ru},
\quad
Ilya L. Shapiro$\,^{c}$\footnote{
E-mail address: ilyashapiro2003@ufjf.br},
\quad
Alexey Toporensky$\,^{d}$\footnote{
E-mail address: atopor@rambler.ru}

\vskip 4mm
\vskip 2mm

{\sl (a) \ Instituto de F\'{\i}sica, UnB,
Campus Universit\'ario Darcy Ribeiro
\\
Bras\'\i lia - DF, 70910-900, Brazil}
\vskip 1mm

{\sl (b) \ Elementary Particles Physics Department,
NRNU MEPhI, Moscow, 115409, Russia}
\vskip 1mm

{\sl (c) \ Departamento de F\'{\i}sica, ICE,
Universidade Federal de Juiz de Fora
\\
Campus Universit\'{a}rio - Juiz de Fora, 36036-900, MG, Brazil}
\vskip 1mm
{\sl (d)
Sternberg Astronomical Institute,
Lomonosov Moscow State University,
\\
Moscow 119991, Russia}
\end{center}

\vskip 6mm

\centerline{\textbf{Abstract}}
\vskip 2mm

\begin{quotation}

\noindent
We consider the background cosmological solutions in the $6D$
(six-dimensional) model with one time and five space coordinates.
The theory of our interest has the action composed by the Einstein
term, cosmological constant, and two conformal terms constructed
from the third powers of the Weyl tensor.  It is shown how the highest
derivative terms in the equations of motion can be isolated that opens
the way for their numerical integration. There are flat anisotropic
solutions which make one of the flat isotropic subspaces to be
static. Depending on the value of bare cosmological constant,
either two-dimensional or three-dimensional subspace can be
static. In particular, there is a physically favorable solution with
three ``large''
space coordinates and two extra inner dimensions stabilized. This
solution is stable for a wide range of coupling constants, but this
requires a special value of the bare cosmological constant.
\vskip 3mm

\vskip 2mm

\end{quotation}

\section{Introduction}
\label{sec1}

Some of the existing approaches to quantum gravity require extra
dimensions. Since the most prospective source of information
about generalized gravity theories is cosmology, there is a
continuous interest in the cosmological solutions with extra
dimensions. In the multidimensional cosmology it is natural to
assume that part of the extra dimensions is compactified.
As the extra dimensional physics is related to quantum effects
and the last are related to high energies, it makes sense to consider
the initial action of multidimensional gravity which may be
consistent in the regime when the masses of the quantum fields
are irrelevant and there is conformal symmetry.
Independent on the version of the fundamental quantum gravity,
at the relatively low (compared to the Planck scale) energy scale
the appropriate description is the quantum theory of matter fields
on curved background. Assuming that the mass spectrum of these
fields is bounded from above at the much lower scale, we arrive
at the problem of the effect of conformal terms in the action of
multidimensional gravity. The purpose of the present work is to
explore cosmological model in the $6D$ spacetime, with four
physical dimensions and two extra space coordinates.

The choice of conformal terms as an addition to the Einstein-Hilbert
term can be regarded as a preparation to explore the $6D$ quantum
corrections derived by integrating the conformal (trace) anomaly.
The integration of trace anomaly \cite{duff77,Duff-94} in $2D$ has
been pioneered by Polyakov \cite{Polyakov81}. The same problem
has been solved in
the 
$4D$ theory by Riegert \cite{rie} and Fradkin and Tseytlin
\cite{FrTs84}. The anomaly and the corresponding effective action
have many important applications. There is an extensive literature
on different aspects of this subject (see, e.g., the reviews
\cite{Duff-94,PoImpo} and the \cite{OUP} for the textbook-level
introduction). It is worth mentioning that the first inflationary
solution has been obtained by Starobinsky \cite{star,star83} for the
model obtained from the trace anomaly \cite{fhh}. Indeed, the
cosmological solutions represent one of the most natural applications
of the anomaly-induced action because it deals with the dynamics of
the conformal factor of the metric.

In $6D$, the integration of trace anomaly has been recently done
\cite{Int6D,Surf6D}, with the important mathematical elements known
from the previous work \cite{Hamada}. Compared to the $4D$ solution,
the result is rather bulky and involves complicated and long
expressions. Thus, a useful starting point for the cosmological
application may be a simplified expression which represents the
classical action of $6D$ gravity compactified to $4D$ for the sake
of physical applications. Such a $6D$  action should provide
renormalizable semiclassical gravity and includes a large amount of
six-derivative terms \cite{Bonora}. However, in the conformal case
renormalizability does not require including all these terms (albeit
their inclusion is not forbidden). Thus, we can restrict the consideration
by the three existing conformal, metric-dependent terms in $6D$
\cite{BFT-2000}.\footnote{The renormalization of these terms and
the topological invariant lead to the main core of anomaly.} Such a
simplification is expected to be useful in getting a general idea of
the complexity of the main problem and the possible ways of its
solution.

The explicit form of the three existing conformal terms is
\cite{BFT-2000}
\begin{align*}
&   C_{\mu\nu\rho\sigma}C^{\mu\al\be\nu}C_{\al\,..\,\be}^{\,\,\;\rho\si}
\\
&   C_{\mu\nu\rho\si}C^{\rho\si\al\be}C_{\al\be\,.\,.}^{\,\,\,\;\;\;\mu\nu}
\\
&   C_{\mu\alpha\beta\gamma}\left( \Box\delta_\nu^\mu
+ 4R^\mu_\nu - \frac{6}{5}R\delta_\nu^\mu\right)
C^{\nu\alpha\beta\gamma }.
\end{align*}
For the sake of simplicity, in this work we consider only the first
two terms. The third term has a different structure from the first
two, representing possible contractions of Weyl tensor. Getting
equations of motion with sixth derivative terms is a complicated
operation, which is different for different structure of terms (see
the Appendices for illustration of this statement) and we choose
to make this reduction. As a basis, we include the standard $6D$
Einstein-Hilbert term with a cosmological constant. Our goal
is to get a solution where extra dimensions are stabilized. It is
known to be difficult getting a stable solution of this type in pure
Einstein gravity (see, e.g., \cite{Petriakova_Sergey} and a particular
example presented below), and, as we will see,  it
is possible to get the desired behavior with the additional conformal
extensions.

The rest of the article is organized as follows. In Sec.~\ref{s2} ,
we describe the conformal $6D$ action and define the line element
describing the $6=4+2$ compactification in the cosmological setting.
Sec.~\ref{s3} considers, as a reference point for subsequent
analysis, the cosmological solution in the pure  $6D$ general
relativity. We assume the $S^2\times S^3$ space section with
initially static $S^2$ and de Sitter expanding $S^3$ space sections.
Sec.~\ref{s4} outlines the derivation and general features of the
field equations in the theory with conformal terms introduced
in Sec.~\ref{s2}, with some additional constraints.
In Sec.~\ref{s5} we consider the solutions with zero spatial
curvature, i.e., with the space section $E ^2\times E^3$, and
obtain stable solutions for this case.
Sec.~\ref{last_sec} concerns the spatially curved case, namely
$S^2\times E^3$, and obtain corresponding stable solutions
for one of the subspaces to be static and another subspace
being inflating.
It is shown that there is the region of the coupling constants
$\theta_1\times \theta_2$ corresponding to the stable static
$S^2$ and to the de Sitter inflating $E^3$.
Finally, in Sec.~\ref{Conc} we draw our conclusions and
describe the possible extensions of the present project.
We follow the conventions that include the signature
$(-,+,+,+)$ in the $4D$ part of the metric, the definition of
the Riemann tensor is
$R_{\,.\,\be\mu\nu}^{\al}=\Ga_{\be\nu,\mu}^{\al}
-\Ga_{\,\be\mu,\nu}^\al...$,
and of the Ricci tensor, $R_{\al\be}=R_{\,.\,\al\mu\be}^\mu$.

\section{Conformal gravity in $6D$}
 \label{s2}

Consider the $6D$ gravitational action
\begin{eqnarray}
&&
S_{g}
\,=\,
\frac{m_4^4}{2}
\int d^6x\sqrt{-g}
\left\{ \frac{R-2\La}{16\pi G}
+ \theta_{1}C_{\mu\nu\rho\sigma}C^{\mu\alpha\beta\nu}
C_{\al\,..\,\beta}^{\,\,\;\rho\sigma}
+ \theta_{2} C_{\mu\nu\rho\sigma}C^{\rho\sigma\alpha\beta}
C_{\alpha\beta..}^{\,\,\;\;\;\mu\nu}\right\},
\mbox{\qquad}
\label{grav.action}
\end{eqnarray}
where
\beq
C_{\alpha\beta\mu\nu}
= R_{\alpha\beta\mu\nu}
+ \frac{R(g_{\alpha\mu}g_{\beta\nu}
- g_{\alpha\nu}g_{\beta\mu})}{(N-2)(N-1)}
- \frac{(R_{\alpha\mu}g_{\beta\nu} - R_{\alpha\nu}g_{\beta\mu}
+ R_{\beta\nu}g_{\alpha\mu} - R_{\beta\mu}g_{\alpha\nu})}{N-2 }
\eeq
is the Weyl tensor with $N=6$, $\La$ is the cosmological constant
and $m_6$ a mass scale which we set to unity. The relation
between the $D$-dimensional Planck mass $m_D$ and the $4D$
one $M_P$ was discussed in \cite{Bronnikov:2023lej} in the
framework of the $f(R)$ gravity. Here we assume $m_D \sim M_P$.

The appropriate line
element for
$R\times S^2\times S^3$ space spacetime is
\begin{align*}
ds^2=-e^{2\alpha}dt^2+e^{2\gamma}\left(d\psi^2+\sin(\psi)^2d\rho^2 \right)
+ e^{2\beta}\left[d\chi^2+\sin(\chi)^2\left(d\theta^2+\sin(\theta)^2d\phi^2
\right) \right]
\end{align*}
though we will consider also limiting cases when one or two subspaces are flat.

The Riemann scalar for the line element (\ref{line.element}) has
a simple form
\beq
R=2e^{-2\alpha}\Big\{ 2\ddot{\gamma}
+ 3\left(\dot{\gamma}\right)^{2}
+ 6\dot{\beta}\dot{\gamma}
- 2\dot{\alpha}\dot{\gamma}
+ 3\ddot{\beta}+6\big(\dot{\beta}\big)^2
- 3\dot{\alpha}\dot{\beta}\Big\}+2e^{-2\gamma}+6e^{-2\beta},
\eeq
where points indicate derivatives with respect to the physical time $t$.

In GR, albeit the Einstein-Hilber action is
second degree in coordinate derivatives, the field equations involve
only second derivatives due to a surface term. Since the line element
\eqref{line.element} is homogeneous, the Weyl scalars in
\eqref{grav.action} are built from the time derivatives of $\al$, $\be$
and $\ga$. The explicit expressions are bulky and hence are postponed
to Appendix \ref{weyl_scalars}. Performing the spatial integrals results
in a volume multiplicative factor and the Lagrangian has a general
structure
\[
L=L(q,\dot{q},\ddot{q}),
\]
where $q = (\alpha,\beta,\gamma)$. The corresponding Euler-Lagrange
equation is
\begin{equation}
\frac{\pa}{\pa q}L
- \frac{d}{dt}\left(\frac{\pa}{\pa\dot{q}}L\right)
+\frac{d^{2}}{dt^{2}}\left(\frac{\pa}{\pa\ddot{q}}L\right)
\,=\,0.
\label{eq.EL}
\end{equation}
Since time does not explicitly enters into the action, the ``energy''
is conserved
\begin{equation}
E=\left(\frac{\pa}{\pa\dot{q}}L\right)\dot{q}
+\left(\frac{\pa}{\pa\ddot{q}}L\right)\ddot{q}
- \frac{d}{dt}\left(\frac{\pa}{\pa\ddot{q}}L\right)\dot{q}
-\frac{d}{dt}\left(\frac{\pa}{\pa\dddot{q}}L\right)\ddot{q}
-L.
\label{eq:constraint}
\end{equation}
The last equation can be also obtained by variation
in $\alpha$ with the subsequent setting $\al = 0$.

Obtaining the dynamical equations for \eqref{grav.action} with the
line element \eqref{line.element} is not completely straightforward
since there are constraints, as shown in Section \ref{s4}, where we
also describe the procedure used for the constrained system. After
this, we assume the existence of de Sitter stages and analytically
explore the corresponding solutions and their linear stability. The
solutions are also investigated numerically for the corresponding
dynamical system. As a numerical check of this procedure, we
verify that the constraints (in each case) are conserved up to the
order of $10^{-10}$, or smaller.

\section{Einstein-Hilbert gravity }
\label{s3}

To construct the reference example, this section presents the
analysis for pure Einstein-Hilbert gravity, i.e.,  we set the constants
$\th_1$ and $\th_2$ in \eqref{grav.action} to zero.
It is known that in the six-dimensional GR with the line element
\eqref{line.element}, there is no solution with  inflating de Sitter
$E^3$ and with a static $E^2$. Also, there is no solution with the
static $E^3$ and exponentially inflating $E^2$, or isotropic, equally
inflating, $E^2$ and $E^3$ spaces (see, e.g., \cite{Rubin2020} and
further references therein).
For this reason, in this section, we consider the spatially curved
$R\times S^2\times S^3$ space-time
as it was done in \cite{Petriakova_Sergey}. The constraint equations
have the form
\beq
&&
\dot{\gamma}^{2}+6\dot{\beta}\dot{\gamma}
+ 3\dot{\beta}^{2} + e^{-2\gamma} + 3e^{-2\beta}
\,=\,\Lambda
\nonumber
\\
&& E = \dot{\gamma}^{2}
+ 6\dot{\beta}\dot{\gamma}
+ 3\dot{\beta}^{2}
+ e^{-2\gamma}
+ 3e^{-2\beta}-\Lambda
= 0
\label{EH00}
\eeq
and the dynamical equations are
\begin{align}
 & \dot{\gamma}^{2} + 3\dot{\beta}\dot{\gamma}
 + \ddot{\gamma}
 + 3\ddot{\beta}
 + 6\dot{\beta}^{2}
 + 3e^{-2\beta}
 =\Lambda
 \nonumber
 \\
 & 3\dot{\gamma}^{2}
 + 4\dot{\beta}\dot{\gamma}
 + 2\ddot{\gamma}
 + 2\ddot{\beta}
 + 3\dot{\beta}^{2}
 + e^{-2\beta}
 + e^{-2\gamma} = \Lambda.
 \label{EH11}
\end{align}

There is an asymptotic 
de Sitter solution
\begin{align}
 & \dot{\beta}=\sqrt{\frac{\Lambda}{6}}=const.\nonumber
 \\
 & \gamma=-\frac{\ln(\Lambda/2)}{2}=const.
 \label{i.cond}
\end{align}
which inflates $S^3$ while $S^2$ remains static.

In the following, we shall illustrate the linearization technique
that is used throughout. First, the highest derivatives are isolated
in  \eqref{EH11} considering the asymptotic $\be\rightarrow\infty$
regime,
\begin{align*}
& \ddot{\beta}=(\Lambda+\dot{\gamma}^2-2\dot{\gamma}\dot{\beta}-9\dot{\beta}^2 +e^{-2\gamma})/4\\
&\ddot{\gamma}=(\Lambda-7\dot{\gamma}^2-6\dot{\gamma}\dot{\beta}+3\dot{\beta}^2 -3e^{-2\gamma})/4.
\end{align*}
Then an equivalent dynamical system is defined
\begin{align*}
&\dot{\beta}=y_1 \\
&\dot{\gamma}=y_2 \\
& \dot{y_1}=(\Lambda+y_2^2-2y_2y_1-9y_1^2 +e^{-2\gamma})/4
\\
&\dot{y_2}=(\Lambda-7y_2^2-6y_2y_1+3y_1^2 -3e^{-2\gamma})/4.
\end{align*}
It is not difficult to check that when $y_1=\sqrt{\Lambda/6}$,
$y_2=0$ and $\gamma=-\ln(\Lambda/2)/2$ the above system
reduces to the solution in \eqref{i.cond} with
$\dot{\beta}=const.,\,\dot{\gamma}=0,\,\dot{y}_1=0$ and
$\dot{y}_2=0$. Expanding near this solution to first order,
gives for the right hand side the equations
\begin{align*}
\left(\begin{array}{c}
\dot{y}_1  \\
\dot{y}_2 \\
\dot{\beta}\\
\dot{\gamma}
\end{array}\right)=
\left( \begin{array}{cccc}
-9\sqrt{\Lambda/6}/2 & -\sqrt{\Lambda/6}/2 & 0 & -\Lambda/4\\
3\sqrt{\Lambda/6}/2 & -3\sqrt{\Lambda/6}/2 & 0 & 3\Lambda/4 \\
1 & 0 & 0 & 0 \\
0 & 1 & 0 & 0
\end{array} \right)
\left(\begin{array}{c}
y_1 \\
y_2\\
\beta\\
\gamma
\end{array}\right)
\end{align*}
The last linear system has the following eigenvalues
\beq
&&
\lambda_{1}=0,
\nn
\\
&&
\lambda_{2}=-\sqrt{3\Lambda/2},
\nn
\\
&&
\lambda_{3}=(-\sqrt{3}+\sqrt{11})\sqrt{\Lambda/8},
\nn
\\
&&
\lambda_{4}=-(\sqrt{3}+\sqrt{11})\sqrt{\Lambda/8}\,.
\label{lambdasEH}
\eeq
Since $\lambda_{3}>0$, the mode associated to this
eigenvalue grows exponentially in time. Without imaginary
part of the eigenvalues, the solution does not show oscillatory
behavior.

\begin{figure}
\begin{center}
\begin{quotation}
\centering
\includegraphics[width=9.0cm,angle=0]{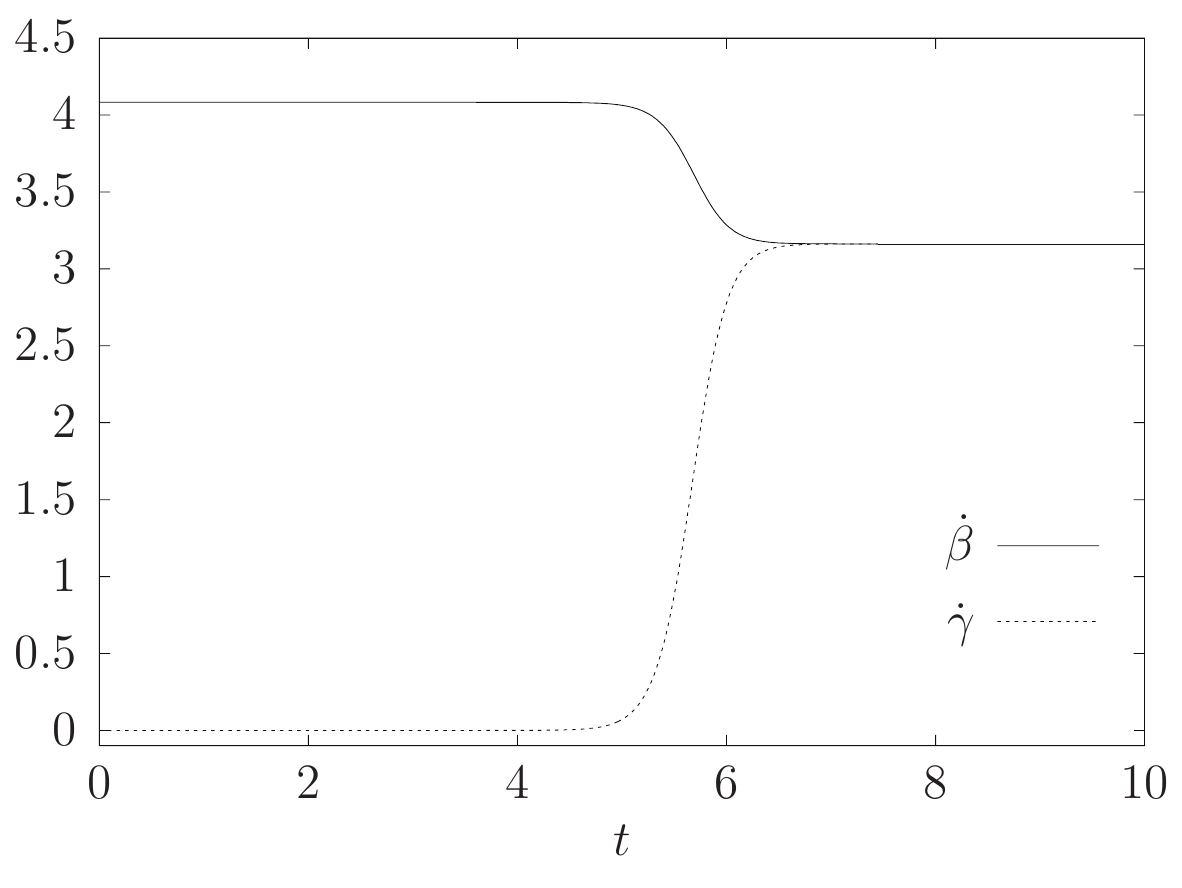}
\caption{\sl
For this plot it is chosen $\Lambda=100$ which results in
$\dot{\beta}_{\mbox{\tiny de Sitter}}\approx 4.08$ given by
(\ref{i.cond}) while $\dot{\gamma}=0$. It can be seen that after
a time interval,  the solution converges as expected to the equally
de Sitter inflating $S^2$ and $S^3$ attractor
$\dot{\be}_{\mbox{{\tiny de Sitter}}}=\dot{\ga}_{\mbox{{\tiny de Sitter}}}
\approx 3.16$ given by (\ref{same.H}).}
\label{Fig1}
\end{quotation}
\end{center}
\end{figure}

There is also an exact de Sitter solution with the same exponential
scale factors for both $S^{3}$ and $S^{2}$
\begin{align}
\dot{\beta}=\dot{\gamma}=\sqrt{\frac{\Lambda}{10}}=const,
\label{same.H}
\end{align}
corresponding to the eigenvalues
\beq
\lambda_{1}
, \quad
\lambda_2  =-\sqrt{\frac{5\Lambda}{2}}
, \quad
 \lambda_3=\lambda_4=0,
\eeq
both with multiplicity 2. This means the de Sitter stage with
equally inflating scale factors is an attractor, as shown in
Fig.~\ref{Fig1}. The initial condition for the orbit in this
figure is $\Lambda=100$ which results in
$\dot{\beta}\approx 4.08$ and $\dot{\gamma}=0$ according
to \eqref{i.cond}. It can be clearly seen in Figure \ref{Fig1}
that after a certain time interval the solution converges to the
equally inflating de Sitter attractor given by \eqref{same.H},
which in this case is
$\dot{\be}_{\mbox{{ \tiny de Sitter} }}
=\dot{\ga}_{\mbox{{ \tiny de Sitter}}}\approx 3.16$.
For the solution plotted in Figure \ref{Fig1} we have explicitly
checked that the constraint $E$ given by \eqref{EH00} fluctuates
with the amplitude about $10^{-12}$ around zero.

The fact that the only stable solution for GR multidimensional
cosmology with a $\Lambda$-term is the de Sitter solution is not
surprising. Below, we see that with the six-derivative Weyl-tensor
dependent terms included, one can induce stable anisotropic
solutions even in the spatially flat case.

\section{Equations of motion in conformal gravity}
\label{s4}

The field equations \eqref{eq.EL}, for the action (\ref{grav.action})
with the line element \eqref{line.element}, with the higher time
derivatives terms separated, can be obtained by taking variations
with respect to the three conformal factors. Taking variation in
$\al$ and dividing the entire equation \eqref{eq.EL} by
$\sqrt{-g}=e^{-\alpha +3\beta +2\gamma}\sin(\psi)\sin(\theta)\sin(\chi)^2$, and then setting
$\alpha=0$, results in equation
\begin{align}
&
E_\alpha=  \frac{9}{50}\left(\dddot{\beta}-\dddot{\gamma}\right)
\biggl\{
\theta_{2}\biggl(-26\dot{\gamma}\ddot{\gamma}
+ 26\dot{\beta}\ddot{\gamma}-34\dot{\gamma}^{3}+8\dot{\beta}^3
+ 26\ddot{\beta}\dot{\gamma} - 50\dot{\beta}^{2}\dot{\gamma}
- 26\dot{\beta}\ddot{\beta}
 \nonumber
\\
&+ 76\dot{\beta}\dot{\gamma}^{2}\biggl) +\left(8 \theta_{2}-18 \theta_{1}\right)\left(\dot{\beta}-\dot{\gamma}\right)e^{-2\gamma}+\left(34 \theta_{2}-\frac{53 \theta_{1}}{2}\right)\left(\dot{\beta}-\dot{\gamma}\right)e^{-2\beta}+\nonumber \\
&
\theta_{1}\bigg(
\frac{17\dot{\gamma}\ddot{\gamma}}{2}
- \frac{17\dot{\beta}\ddot{\gamma}}{2}
+ \frac{53\dot{\gamma}^{3}}{2}-71\dot{\beta}\dot{\gamma}^{2}
- \frac{17\ddot{\beta}\dot{\gamma}}{2}
+ \frac{125\dot{\beta}^{2}\dot{\gamma}}{2}
+ \frac{17\dot{\beta}\ddot{\beta}}{2}
- 18\dot{\beta}^{3}\bigg)\bigg\}+F_{\alpha}.
\label{Ealpha}
\end{align}
Using similar procedures for variations in $\beta$ and $\gamma$
result in the equations
\begin{align}
& E_\beta=\left(\ddddot{\gamma}-\ddddot{\beta}\right)
\bigg\{ \frac{9 \theta_{1}}{100}\biggl(17\ddot{\gamma}
+ 53\dot{\gamma}^{2}-89\dot{\beta}\dot{\gamma}
- 17\ddot{\beta}+36\dot{\beta}^{2}+36e^{-2\gamma}+53e^{-2\beta}\biggr)
 \nonumber
 \\
&
\qquad 
- \frac{9 \theta_{2}}{25}\Bigl(13\ddot{\gamma}
 +17\dot{\gamma}^{2} - 21\dot{\beta}\dot{\gamma}-13\ddot{\beta}
+ 4\dot{\beta}^{2}+4e^{-2\gamma}+17e^{-2\beta}\Bigl)\biggr\}+F_\beta,
\label{Ebeta}
\end{align}
\begin{align}
 &
 E_\gamma
 = \left(\ddddot{\beta}-\ddddot{\gamma}\right)
 \bigg\{
 \frac{9\theta_{1}}{100}\biggl(17\ddot{\gamma}+53\dot{\gamma}^{2}
 -89\dot{\beta}\dot{\gamma}-17\ddot{\beta}+36\dot{\beta}^{2}+53e^{-2\beta}+36e^{-2\gamma}\biggr)
 \nonumber
\\
&
\qquad 
- \frac{9 \theta_{2}}{25}\biggl(13\ddot{\gamma}
+ 17\dot{\gamma}^{2}
- 21\dot{\beta}\dot{\gamma}-13\ddot{\beta}
+ 4\dot{\beta}^{2}+4e^{-2\gamma}+17e^{-2\beta}\biggr)\biggr\}+F_\gamma\,.
\label{Egamma}
\end{align}
The bulky expressions for $F_\alpha$, $F_\beta$ and $F_\gamma$
can be found in Appendices \eqref{F_alpha},  \eqref{F_beta}, and
\eqref{F_gamma}, respectively.

Usually, the desired dynamics is obtained directly by setting equations
\eqref{Ealpha} - \eqref{Egamma} to zero $E_\alpha=0$, $E_\beta=0$
and $E_\gamma=0$. However, it is easily seen that it is not possible to
isolate the higher derivatives in \eqref{Ebeta} and \eqref{Egamma}.
Besides that, as already mentioned, \eqref{Ealpha} is a constraint
which must be preserved in time.

The conformal invariant part of the gravitational equations does
not contribute to the trace of equations of motion. This, as should
be expected, its form is rather simple
\beq
E_{\alpha+\beta+\gamma}
=E_\alpha+E_\beta+E_\gamma
=\ddot{\gamma}
+\frac{3}{2}\ddot{\beta}
+\frac{3}{2}\dot{\gamma}^{2}
+3\dot{\beta}\dot{\gamma}
+3\dot{\beta}^{2}+\frac{1}{2}e^{-2\gamma}+\frac{3}{2}e^{-2\beta}-\frac{3\Lambda}{4}=0.
\label{eq:trace}
\eeq
It is easy to see that the trace equation is a constraint in the dynamical
system since it does not involve the higher time derivatives.

As a next step, let us derive an identity which will be useful in
the following. Consider left invariant 1-form basis for $S^2$ and $S^3$
\beq
ds^2 = - e^{2\alpha}dt^2 + e^{2\gamma} \left[
(d\omega^1)^2+(d\omega^2)^2 \right]+e^{2\beta}\left[(d\alpha^1)^2+(d\alpha^3)^2+(d\alpha^3)^2\right].
\eeq
With this choice of the base, the spatial part of the connection
follows from the commutators of the tetrads. The temporal part
of the connection is given by the Christoffel symbols, since the
base vectors do not depend on time and the commutators which
involve time are zeroes. Thus, the temporal, non-zero components of the
connection are $(\alpha=0)$
\begin{align}
&\Gamma^i_{0i}=\left\{ \begin{array}{l l}
\dot{\beta} & \mbox{for $i=1,2,3$
\quad Sphere sector $S^3$}
\\
\dot{\gamma} & \mbox{for $i=4,5$
\quad  \quad  Sphere sector $S^2$}
\end{array}\right. \nonumber \\
&
\Gamma^0_{ii}=\left\{ \begin{array}{l l}
\dot{\beta}e^{2\beta} & \mbox{for $i=1,2,3$ \quad
Sphere sector $S^3$}
\\
\dot{\gamma}e^{2\gamma} & \mbox{for $i=4,5$
\quad \quad Sphere sector $S^2$}
\end{array}\right.
\label{Gamma}
\end{align}

Consider the variational derivatives of the action
\eqref{grav.action}
\beq
&&
E_\alpha=\frac{1}{\sqrt{-g}}\Bigg\{ \frac{\delta S_g}{\delta \alpha}
= \frac{\partial g_{00}}{\partial \alpha}
\frac{\delta S_g}{\delta g_{00}} =
- 2 e^{2\alpha} \frac{\delta S_g}{\delta g_{00}}\Bigg\} \,,
\nonumber
\\
&&
E_\gamma=\frac{1}{\sqrt{-g}}\Bigg\{ \frac{\delta S_g}{\delta \gamma}=\left[
\frac{\partial g_{44}}{\partial \gamma}
\frac{\delta S_g}{\delta g_{44}}
+ \frac{\partial g_{55}}{\partial \gamma}
\frac{\delta S_g}{\delta g_{55}}\right]
=4e^{2\gamma}\frac{\delta S_g}{\delta g_{55}}
=4e^{2\gamma}\frac{\delta S_g}{\delta g_{44}} \Bigg\}\,,
\nonumber
\\
&&
E_\beta=\frac{1}{\sqrt{-g}} \Bigg\{ \frac{\delta S_g}{\delta \beta}
=\left[ \frac{\partial g_{11}}{\partial \beta}
\frac{\delta S_g}{\delta g_{11}}
+ \frac{\partial g_{22}}{\partial \beta}
\frac{\delta S_g}{\delta g_{22}}
+\frac{\partial g_{33}}{\partial \beta}
\frac{\delta S_g}{\delta g_{33}}\right]
\nonumber
\\
&&
\qquad \quad
= 6e^{2\beta}\frac{\delta S_g}{\delta g_{11}}
= 6e^{2\beta}\frac{\delta S_g}{\delta g_{22}}
= 6e^{2\beta}\frac{\delta S_g}{\delta g_{33}}\Bigg\}\,.
\label{E-EL,EC}
\eeq
If both sides of each of the above equations is divided by
$\sqrt{-g}=e^{2\gamma+3\beta}$ ($\alpha=0$), the
right hand sides of \eqref{Ealpha}-\eqref{Egamma} give
the components of the field tensor
\[
E^{ab}=\frac{1}{\sqrt{-g}}\frac{\delta}{\delta g_{ab}}S_g,
\]
which satisfy $\nabla_a E^{ab}=0$. Now \eqref{E-EL,EC}
establishes the equality between \eqref{Ealpha}-\eqref{Egamma}
and the components of $E^{ab}$  ($\alpha=0$),
\beq
&&
E_\alpha=-2 E^{00},
\qquad
E_\gamma=4e^{2\gamma}E^{44} = 4e^{2\gamma}E^{55}\,,
\nn
\\
&&
E_\beta=6e^{2\beta}E^{11}
= 6e^{2\beta}E^{22}= 6e^{2\beta}E^{33}.
\label{Etensor}
\eeq
As tensor $E^{ab}$ in \eqref{Etensor} satisfies
\beq
\nabla_a E^{ab}=0
\quad
\Longrightarrow
\quad
\partial_0E^{0b}+\Gamma^a_{na}E^{nb}
+ \Gamma^b_{an}E^{an} = 0,
\eeq
considering
the connection \eqref{Gamma} we get, for for $b=0$,
\beq
&&
-\frac{1}{2}\partial_t E_\alpha
- \frac{1}{2}\Gamma^a_{0a}E_\alpha+\Gamma^0_{ii}E^{ii}=0,
\nonumber
\\
&&
- \,\partial_t E_\alpha
- (2\dot{\gamma} + 3\dot{\beta})E_\alpha
+ 2 \dot{\beta}e^{2\beta}3E^{11}
+ 2 \dot{\gamma}e^{2\gamma}2E^{44}= 0 ,
\nonumber
\\
&&
-\,\partial_t E_\alpha - (2\dot{\gamma}
+ 3\dot{\beta})E_\alpha
+ \dot{\beta}E_\beta
+ \dot{\gamma}E_\gamma \equiv 0\,.
\label{cov_div}
\end{eqnarray}
This last identity can be verified by direct replacement of
the equations of motion \eqref{Ealpha}-\eqref{Egamma}
in \eqref{cov_div}.

To obtain a consistent dynamical system we proceed as follows.
The time derivative of the trace equation \eqref{eq:trace},
$\dot{E}_{\alpha+\beta+\gamma}$, and the ``energy'' equation
\eqref{Ealpha}, $E_\alpha=0$ are both set to zero, namely
\beq
\dddot{\gamma}+\frac{3}{2}\dddot{\beta}
+ 3\dot{\gamma}\ddot{\gamma}
+ 3\ddot{\beta}\dot{\gamma}
+ 3\dot{\beta}\ddot{\gamma}
+ 6\dot{\beta}\ddot{\beta}-\dot{\gamma}e^{-2\gamma}-3\dot{\beta}e^{-2\beta}=0,\label{dddbetadddgamma}
\eeq
and
\beq
&&\frac{9}{50}\left(\dddot{\beta}-\dddot{\gamma}\right)
\biggl\{
\theta_{2}\biggl(-26\dot{\gamma}\ddot{\gamma}
+ 26\dot{\beta}\ddot{\gamma}-34\dot{\gamma}^{3}+8\dot{\beta}^3
+ 26\ddot{\beta}\dot{\gamma} - 50\dot{\beta}^{2}\dot{\gamma}
- 26\dot{\beta}\ddot{\beta}
 \nonumber
\\
&&+ 76\dot{\beta}\dot{\gamma}^{2}\biggl) +\left(8 \theta_{2}-18 \theta_{1}\right)\left(\dot{\beta}-\dot{\gamma}\right)e^{-2\gamma}+\left(34 \theta_{2}-\frac{53 \theta_{1}}{2}\right)\left(\dot{\beta}-\dot{\gamma}\right)e^{-2\beta}+\nonumber \\
&&
\theta_{1}\bigg(
\frac{17\dot{\gamma}\ddot{\gamma}}{2}
- \frac{17\dot{\beta}\ddot{\gamma}}{2}
+ \frac{53\dot{\gamma}^{3}}{2}-71\dot{\beta}\dot{\gamma}^{2}
- \frac{17\ddot{\beta}\dot{\gamma}}{2}
+ \frac{125\dot{\beta}^{2}\dot{\gamma}}{2}
+ \frac{17\dot{\beta}\ddot{\beta}}{2}
- 18\dot{\beta}^{3}\bigg)\bigg\}\nonumber\\
&&+F_{\alpha}=0.
 \label{dyn_sys}
\eeq
It is possible to isolate the higher derivatives
$\dddot{\beta}$ and $\dddot{\gamma}$ from
\eqref{dddbetadddgamma} and \eqref{dyn_sys},
resulting in a dynamical system. The initial conditions for this system
must be defined to be consistent with the original Euler-Lagrange
equations \eqref{Ebeta} and with \eqref{Egamma}, i.e.,
$E_{\alpha+\beta+\gamma}=0$, at the initial point. A direct
consequence of \eqref{eq:trace} and \eqref{cov_div} when
$E_\alpha=0$ and $E_{\alpha+\beta+\gamma}=0$,  is that both
$E_\beta$ and $E_\gamma$ are also zero, $E_\beta=0$ and
$E_\gamma=0$. Besides that, we have explicitly checked that
the time derivative of both equations in \eqref{dyn_sys}, when
substituted into the initial Euler-Lagrange equations \eqref{Ebeta}
and \eqref{Egamma}, result in $E_\beta=0$ and $E_\gamma=0$,
respectively.

\section{Spatially $E^3\times E^2$ anisotropic solutions}
\label{s5}

This section specializes to spatially flat case when the metric can be
written in the form
\begin{equation}
ds^{2}=-e^{2\alpha(t)}dt^{2}+e^{2\gamma(t)}(du^2+dv^2)
+e^{2\beta(t)}(dx^2+dy^2+dz^2)\,.
\label{line.element}
\end{equation}
Here $x$, $y,$ and $z$ are the coordinates in $E^{3}$ and $u$ and $v$
are coordinates in $E^{2}.$

Inflationary de Sitter type solutions are obtained in the limit when both $\beta\rightarrow \infty$ and $\gamma\rightarrow \infty$ into the dynamical
system defined in \eqref{dddbetadddgamma} and \eqref{dyn_sys} subject to the constraint
\eqref{eq:trace}. For this, the derivatives of order 3, which are the
highest ones in \eqref{dyn_sys}, \eqref{dddbetadddgamma} must be zero. Besides this the derivatives of order 2 must also be zero and the initial
condition is $\dot{\beta}=H=const.$, $\dot{\gamma}=0$,
$\ddot{\beta}=0$ and $\ddot{\gamma}=0$. Then, the dynamical system \eqref{dyn_sys}, \eqref{dddbetadddgamma} subject to the initial constraint \eqref{eq:trace} is drastically simplified to
\begin{align*}
 & \Lambda-4H^2=0 \\
&\frac{H^2}{2}+\frac{39H^6\theta_2}{50}-\frac{51H^6\theta_1}{200}=0,
\end{align*}
which results in the de Sitter expanding $E^3$ solution, with static $E^2$
\begin{align}
&\dot{\beta}=H=\frac{2\sqrt{5}}{[3(17 \theta_1-52 \theta_2) ]^{1/4} }\nonumber\\
 & \dot{\gamma}=0\nonumber \\
 &\Lambda=\frac{80}{\sqrt{3}\sqrt{17 \theta_1-52 \theta_2}}\label{de SitterC3}
\end{align}
which inflates $E^3$ while maintains a static $E^2$, where $\theta_1$ and $\theta_2$ are coefficients in \eqref{grav.action}.

\begin{figure}[htpb]
\begin{centering}
 \resizebox{\imsize}{!}{\includegraphics{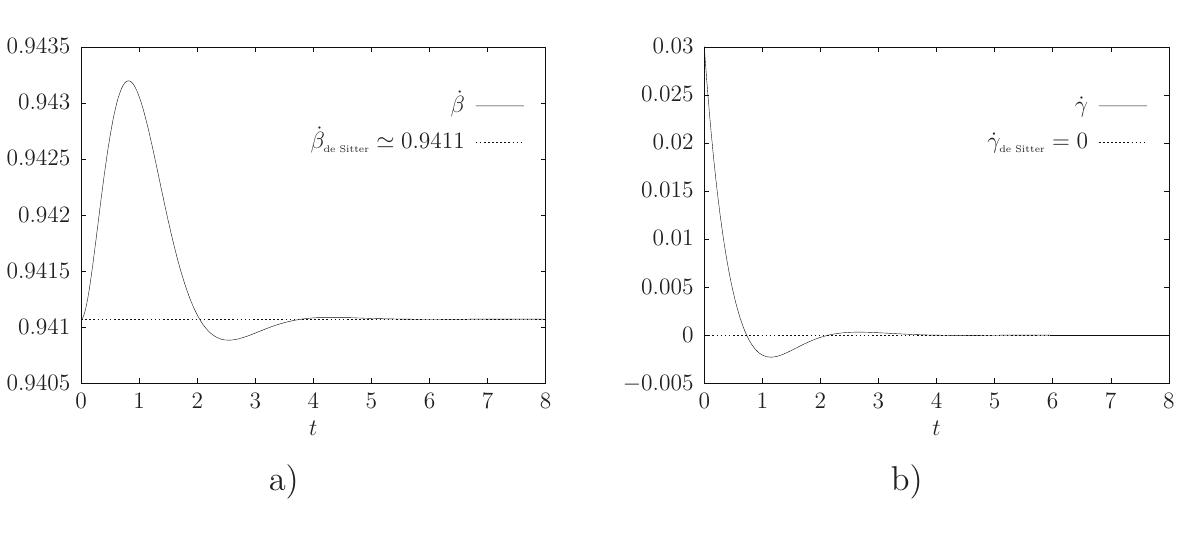}}
 \tabularnewline
\par\end{centering}
\caption{\sl
In this plot the choice of coupling constants is  $ \theta_{1}=-2.235$
and $ \theta_{2}=-4$. The initial condition is not chosen exactly as
the de Sitter solution,  \eqref{de SitterC3} for
$\dot{\beta}_{\mbox{{\tiny de Sitter}}}\simeq 0.9411$, while
$\dot{\gamma}=0$. The initial condition for the solution plotted
as a solid line is chosen as
$\dot{\beta}=\dot{\beta}_{\mbox{{\tiny de Sitter}}}$ and
$\dot{\gamma}=0+0.03$. In panel a) it is shown that
$\dot{\beta}\rightarrow \dot{\beta}_{\mbox{{\tiny de Sitter}}}
\simeq0.9411$
the inflating $E^3$ asymptotically,  while in panel b) it is shown that
$\dot{\gamma}\rightarrow 0$ the static $E^2$ also asymptotically.}
\label{f1}
\end{figure}

In Figure \ref{f1} the plots shown in dotted lines are with respect to the specific choices of $ \theta_{1}=-2.235$ and  $ \theta_{2}=-4$ 
in the exact solution \eqref{de SitterC3}, which have static Hubble parameters namely, $\dot{\beta}_{\mbox{\tiny{de Sitter}}}=0.9411$ and $\dot{\gamma}=0$. Figure \ref{f1} also shows plotted as a solid line, a solution very near this static solution with $\dot{\beta}$ and $\Lambda$ given by \eqref{de SitterC3} while $\dot{\gamma}$ is chosen as $\dot{\gamma}=0+0.03$ instead of $\dot{\gamma}=0$.

 For the specific choice  $ \theta_{1}=-2.235$ and $ \theta_{2}=-4$ 
 made in Figure \ref{f1} the linearization of the system \eqref{dyn_sys} give the following eigenvalues
\begin{align}
&&\lambda_1\simeq-1.41-1.79 i && \lambda_2\simeq-1.41+1.79 i && \lambda_3\simeq -2.82 &&\lambda_4=0. \label{lambdas}
\end{align}
We see that all $Re\{\lambda_i\}< 0$. Figure \ref{f1} shows, in accordance to linear analysis, that the deviation from the exact static solution, \eqref{de SitterC3}, fluctuate and are exponentially damped  such that the solution asymptotes the de Sitter static solution. We have numerically tested the possible combinations of initial deviations sufficiently near this exact solution in \eqref{de SitterC3}, from above and bellow in $\dot{\beta}$ and $\dot{\gamma}$, and confirmed that asymptotically the solution converges to  \eqref{de SitterC3}. Thus, both the above linear analysis and the numerical solutions show that de Sitter expanding solution for $E^3$ for static $E^2$ given in  \eqref{de SitterC3} is an attractor for this choice of coupling constants.

\begin{figure}[htpb]
\begin{centering}
 \resizebox{\halfsize}{!}{\includegraphics{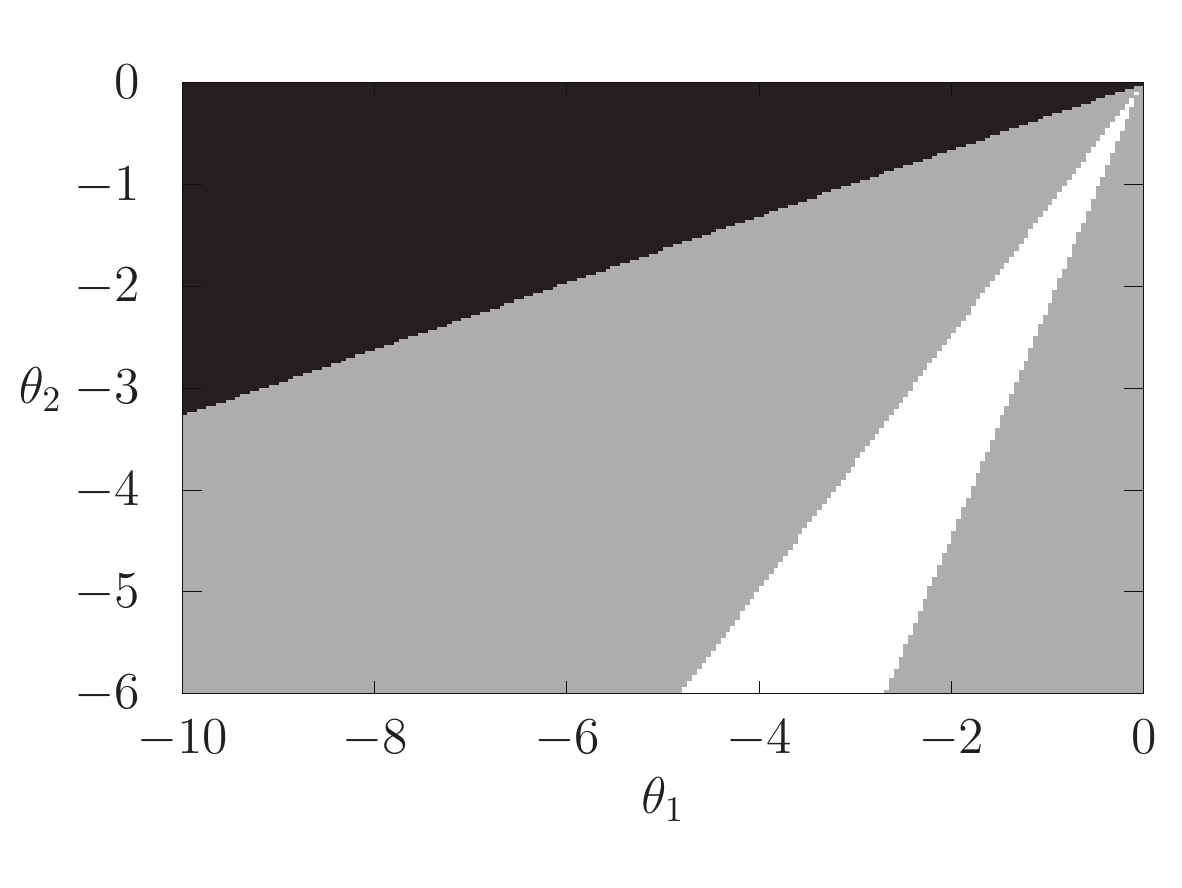}}
  \tabularnewline
 \tabularnewline
\par\end{centering}
\caption{\sl This plot refers to the stability of the solution
\eqref{de SitterC3} with a de Sitter inflating $E^3$ with a
static $E^2$ in the coupling constants $\theta_1 \times \theta_2$
plane. For every white point all the eigenvalues have negative real
part. While for every gray point, at least one eigenvalue has
positive real part. The black points represent the region for
which the solution \eqref{de SitterC3} is imaginary, is not
physically relevant.
\label{theta_stable_region}}
\end{figure}

Figure \ref{theta_stable_region}, shows a grid of $200\times 200$
values for the coupling constants $\theta_1\times \theta_2$.
The tones:
black, gray or white,  attributed to each point of the grid, reflects
the linearized eigenvalues with respect to the solution shown in
\eqref{de SitterC3}, for inflating $E^3$ and static $E^2$. In Figure
\ref{theta_stable_region} every white point corresponds to a specific
choice of $\theta$s for which all the real parts of the eigenvalues
are negative, as in \eqref{lambdas}. For example,  the specific
orbit shown in Figure \ref{f1} which results in these $\lambda$'s
\eqref{lambdas}, namely $\theta_1\simeq -2.24$ and $\theta_2=-4$:
it can be seen in Figure \ref{theta_stable_region} that this orbit
corresponds to a white point. Black points in
Figure~\ref{theta_stable_region} correspond to the region where
$52\theta_2>17\theta_1$ and the solution \eqref{de SitterC3}
becomes imaginary. Gray points are values for $\theta_1$ and
$\theta_2$, which at least one of the real part of the linearized
eigenvalues concerning the solution shown in \eqref{de SitterC3}
is positive showing its instability. The white region is the stable
region for the solution of expanding $E^3$ with static $E^2$.

It is also possible to obtain a solution which inflates $E^2$ while
keeping a static $E^3$
\begin{align}
&\dot{\beta}=0\nonumber\\
 & \dot{\gamma}=\frac{2\sqrt{5}}{[6(9 \theta_1-4 \theta_2) ]^{1/4}}
 \nonumber
 \\
 &\Lambda=\frac{20}{3 \sqrt{9 \theta_1-4 \theta_2}}
 \label{de SitterC3_S2}.
\end{align}
Let us remember that Newton constant is set to $G=1/(16\pi)$.

\begin{figure}[htpb]
\begin{centering}
\resizebox{\imsize}{!}{\includegraphics{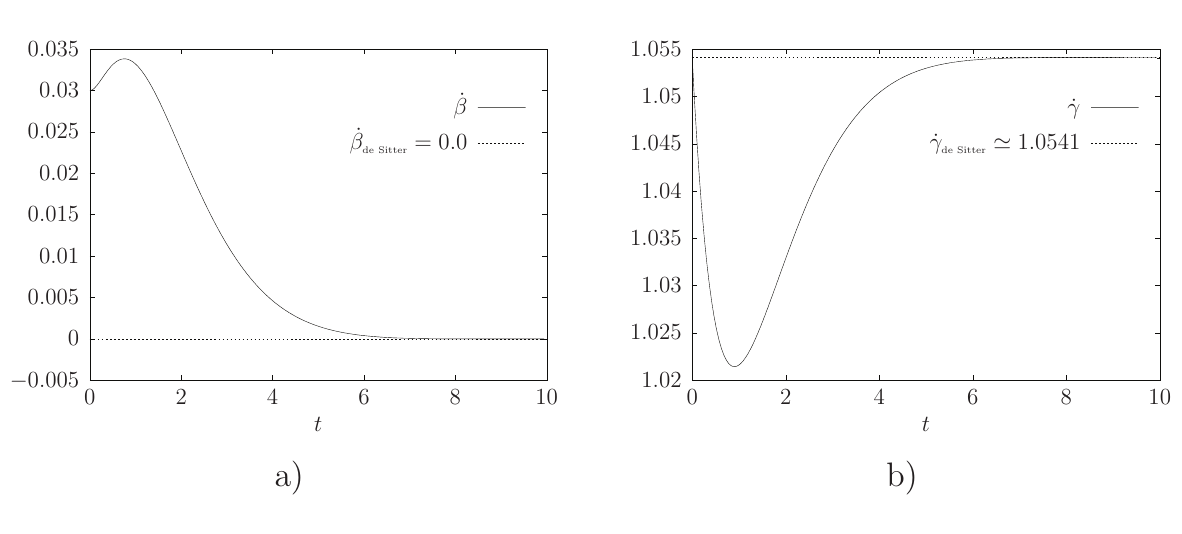}}
\tabularnewline
 \tabularnewline
\par\end{centering}
\caption{\sl For this plot it is chosen $ \theta_{1}=1.88$,
$ \theta_{2}=2$ and the de Sitter inflating $E^2$  specified
by \eqref{de SitterC3_S2} results in
$\dot{\gamma}_{\mbox{{\tiny de Sitter}}}\simeq 1.0541$
with static $E^3$, $\dot{\beta}=0$. The solid line in the plot
shows an orbit with initial condition near this solution with
$\dot{\gamma} =\dot{\gamma}_{\mbox{{\tiny de Sitter}}}$
and $\dot{\beta}=0+0.02$. Panel a) shows that the orbit approaches $\dot{\gamma}\rightarrow\dot{\gamma}_{\mbox{{\tiny de Sitter}}}$
asymptotically while panel b) shows that $\dot{\beta}\rightarrow 0$
asymptotically.}
\label{f2}
\end{figure}

The linearized eigenvalues for the system \eqref{dyn_sys} shown
in Figure \ref{f2} have real part $Re\{\lambda_i\}\leq 0$ such that
this solution is also an attractor
\begin{align}
&&  \lambda_1\simeq-1.05+0.480 i &&\lambda_2\simeq-2.11
&&   \lambda_3=-1.05-0.480 i
&&     \lambda_4\simeq 0.
\label{lambda S2}
\end{align}

The  dotted lines in Fig.~\ref{f2} show the de Sitter stage
\eqref{de SitterC3_S2}, with the choice $\theta_{1}=1.88$ and
$\theta_{2}=2$ which results in inflating $E^2$
$\dot{\gamma}_{\mbox{\tiny{de Sitter}}} \simeq 1.0541$ and
static $E^3$. It is also shown in Figure \ref{f2} that an initial
condition near this solution
$\dot{\gamma}_{\mbox{\tiny{de Sitter}}} \simeq1.0541$ with
$\dot{\beta}=0$ approaches the solution \eqref{de SitterC3_S2},
$\dot{\beta}\rightarrow 0$ and
$\dot{\gamma}\rightarrow \dot{\gamma}_{\mbox{\tiny{de Sitter}}}$.
The solution \eqref{de SitterC3_S2} is an attractor for this choice
of coupling constants, also in accordance to the linearized analysis
\eqref{lambda S2}.
Concerning the solution presented in Figure \ref{f2}, we have
checked that the constraint $E_{\alpha+\beta+\gamma}$ given
in \eqref{eq:trace} fluctuates with the amplitude about
$10^{-9}$ around zero.

We see that in order to get a solution with extra dimension
stabilization, the bare cosmological constant $\Lambda$ should
take a particular value determined by coupling constant. In the next
section we will see that non-zero spatial curvature allows us to lift
this restriction.

\section{Spatially $E^3\times S^2$ solutions}
\label{last_sec}

In this section,  we demonstrate that extra dimensions can be
compact and sufficiently small to avoid detection. Consider a
space-time with compact inner space $S^2$, $R\times E^3\times S^2$.
Similar to the previous section, the initial conditions are chosen
as $\beta\rightarrow \infty$, $\dot{\beta}_{\mbox{\tiny{de Sitter}}}=H=const$ and
$\gamma=const$ into \eqref{dddbetadddgamma}, \eqref{dyn_sys} and constraint \eqref{eq:trace}.
\begin{align}
&
\gamma=\ln \bigg(
\frac{\sqrt{2}}{\sqrt{3}\sqrt{\Lambda-4H^2}}\bigg),
\nonumber
\\
& \theta_{1}={{2106\left(2\theta_{2}\Lambda^3-22H^2
 \theta_{2}\Lambda^2+80H^4\theta_{2}\Lambda\right)
 +50\left(
 16(\Lambda-6H)-4056H^6\theta_{2}\right)}\over{51
 \left(3\Lambda-13H^2\right)\left(3\Lambda-10
 H^2\right)^2}},
 \label{sol:deSitterE3staticS2}
\end{align}
which describes the $4D$ de Sitter space with a static $S^2$ section.
Fixing $\Lambda$, $\theta_1$ and $\theta_2$ allows us to obtain the
Hubble parameter $H$.  According to the first line in
Eq.~\eqref{sol:deSitterE3staticS2}, the extra space size $e^{\ga}$
is large compared to the Planck scale for the specific $H$, validating
the classical equations used above.

\begin{figure}[htpb]
\begin{centering}
 \resizebox{\imsize}{!}{\includegraphics{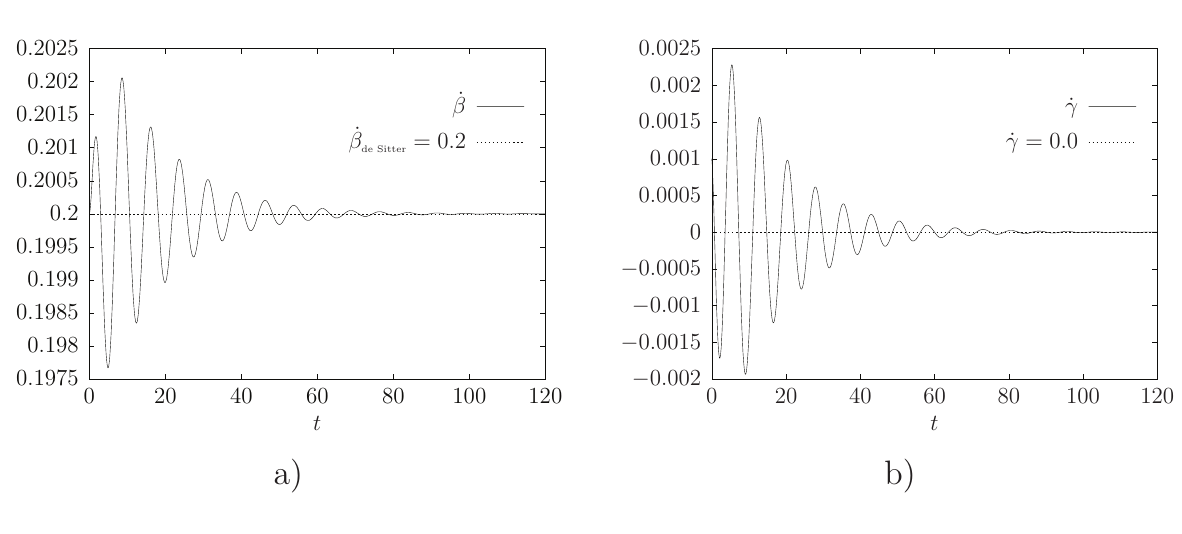}}
\tabularnewline
 \tabularnewline
\par
\end{centering}
\caption{\sl For the dotted plots, the initial condition is chosen as
$\theta_{2}=-0.5$ and $\Lambda=0.6$ and the de Sitter space
$R\times E^3$, $\dot{\beta}_{\mbox{{\tiny de Sitter}}}=0.2$
with static $S^2$, $\dot{\gamma}=0$ which according to
 \eqref{sol:deSitterE3staticS2} results in
$\gamma_{\mbox{\tiny{$S ^2$}} } \simeq 1.231$
 and $\theta_1\simeq 0.721$. The solid lines show an orbit with initial
 condition near this doted line solution, with the only difference that
 $\dot{\gamma} =0+0.001$ instead of $\dot{\gamma}=0$. Panel
 a) shows that the orbit approaches
 $\dot{\beta}\rightarrow\dot{\beta}_{\mbox{{\tiny de Sitter}}}$
 asymptotically, while panel
b) shows that $\dot{\gamma}\rightarrow 0$ asymptotically.}
\label{f3}
\end{figure}

\begin{figure}[htpb]
\begin{centering}
 \resizebox{\imsize}{!}{\includegraphics{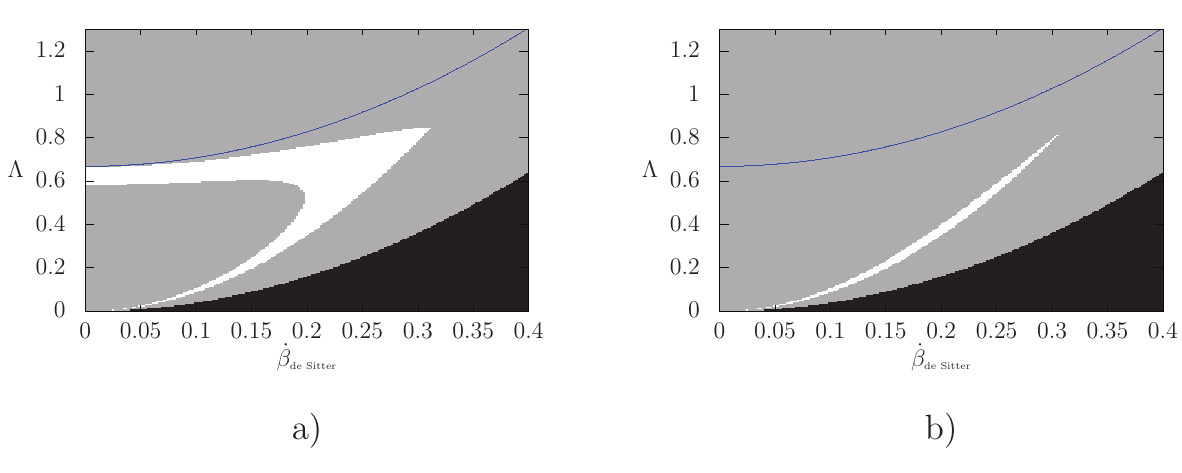}}
 \tabularnewline
\par\end{centering}
\caption{\sl
This plot considers stability of the solution in \eqref{sol:deSitterE3staticS2}.
It shows a grid of $200\times200$ values for  $\Lambda$ and initial Hubble
parameter $\dot{\beta}_{\mbox{\tiny{de Sitter}}}$ considering the solution \eqref{sol:deSitterE3staticS2}.
For every white point the real parts of all of the eigenvalues are negative.
For gray points ,at least one of the real part of the eigenvalues is positive.
Above and bellow the blue line, the radius of the additional space $S^2$,
$e^\gamma$, is respectively smaller and bigger than Planck unit. The
boundary gray/black corresponds to $\gamma\rightarrow \infty$ and the
black points exclude the particular de Sitter solution given by
\eqref{sol:deSitterE3staticS2}, because this solution requires
$\Lambda\geq 4H^2$. Panel a) $\theta_2=-0.5$ and in panel
b) $\theta_2=0.5$. It can be seen in panel a) that negative $\theta_2$
allows stable solutions very close to spatially static space times.}
 \label{f4}
\end{figure}

The solution \eqref{sol:deSitterE3staticS2} is shown as a dotted line
in Figure \ref{f3}, while the solid line shows a solution sufficiently
close to it that is attracted to \eqref{sol:deSitterE3staticS2}
asymptotically. The initial condition chosen for the solution plotted
as the solid line is  $\theta_{2}=-0.5$ and $\Lambda=0.6$ with
$\dot{\beta}_{\mbox{{\tiny de Sitter}}}=0.2$ and
$\dot{\gamma}=0.001$ instead of $\dot{\gamma}=0$ which
asymptotically approaches  \eqref{sol:deSitterE3staticS2} with
$\gamma_{\mbox{\tiny{$S ^2$}}}\simeq 1.231$,
$\theta_1\simeq 0.721$ and this solution has the following
eigenvalues
\beq
&&
\lambda_1\simeq-0.538-0.835 i,
\qquad
\lambda_2\simeq-0.0619-0.835 i,
\qquad
\lambda_3= -0.538+0.835 i \,,
\nonumber
\\
&&
\lambda_4 \simeq -0.0619+0.835 i,
\qquad
\lambda_5=0.
\eeq
Linear analysis is in accordance, and indicates that the solution
\eqref{sol:deSitterE3staticS2} is an attractor for the choices of
$\theta_{2}=-0.5$, $\Lambda=0.6$ and
$\dot{\beta}_{\mbox{{\tiny de Sitter}}}=0.2$.

As in the previous section, Figure \ref{f4}
 shows a grid of $200\times 200$ values for
 $H=\dot{\beta}_{\mbox{\tiny{de Sitter}}} \times \Lambda$.
 The white regions in Figure \ref{f4} show the stability regions
for the solutions in \eqref{sol:deSitterE3staticS2}, all eigenvalues
have a negative real part. The gray region is unstable, at least one
of the real parts of the eigenvalues is positive, while the black
region excludes the solution \eqref{sol:deSitterE3staticS2}, since
\eqref{sol:deSitterE3staticS2}
requires $\Lambda\geq 4H^2$. The boundary gray/black occurs when
$\Lambda=4H^2$ for which the radius of the inner space $S^2$ goes
to infinity, $e^\gamma \rightarrow \infty$. The blue line divides
regions in which the radius of $S^2$, is bigger and smaller than
one Planck unit. Above the blue line $S^2$ has radius smaller
than unity, and bellow the blue line $S^2$ has radius bigger than
unity.
The left panel a), for negative $\theta_2$ reflects the behavior of
the stable white region with the presence of a plateau for several
values of $\dot{\beta}_{\mbox{\tiny{de Sitter}}}$ with a narrow
range of values for  the cosmological constant $\Lambda$. It is
also visual from Figure \ref{f4} that it is possible a stable
situation even for both static $E^3$ and static $S^2$, because
the plateau allows the white region with arbitrary small
$\dot{\beta}_{\mbox{\tiny{de Sitter}}}$.  As $\theta_2$ is
decreased more negative then the value showed in  panel a) of
Fig.~\ref{f4}, the plateau is also decreased to smaller values of
$\Lambda$. Panel b) reflects also the behavior of the white region
when $\theta_2>0$ with the absence of the plateau. For increasing
values of $\theta_2$ passed the value showed in panel b), the white
stripe moves uppwards and to the left. Since there is no plateau,
panel b) of Fig.~\ref{f4} shows that it is not possible to obtain
arbitrarily slow expanding $E^3$ with static $S^2$ when $\th_2 > 0$.

\section{Conclusions}
\label{Conc}

 In the present paper, we considered cosmological consequences
 of including the two Weyl (conformal) invariant terms in six
 spacetime dimensions. First of all, we have shown how to isolate
 highest derivative terms in the equations of motion, so that these
 equations can be integrated numerically. Then, we present flat
 anisotropic solutions which make one of the flat isotropic
subspaces to be static. In our $5+1$ dimension setting, depending
on the value of the bare cosmological constant, either 2-dimensional
or 3-dimensional subspace can be static. Naturally, the case of
expanding 3-dim with static 2-dim spaces is more interesting
since it represents a solution with big ``our'' world and the
stabilization of inner dimensions. We have shown that this solution
is stable for a wide range of coupling constants of the theory under
investigation. A relevant detail is that such a solution requires a
fixed particular value of the cosmological constant.

The required value of the  cosmological constant is very large 
compared to the magnitude of the  cosmological constant which is 
observed in our four-dimensional Universe. There is no contradiction
in this difference because the extra dimensions we are discussing 
are assumed to become invisible in the very early epoch, which 
means before the electroweak phase transition. Thus, the value
of the overall (observed) cosmological constant should be much 
greater that the result of the fine tuning between vacuum and 
induced counterparts we meet in the nowadays Universe 
\cite{Weinberg89}.

A solution where positively curved two-dimensional subspace is 
static could be found in all cases. For such a solution it is 
not necessary to
fix the value of cosmological constant. Again, a solution of this
type can be stable for a non-zero measure region in  the solution
space. We can mention a similarity with the corresponding
types of solutions in four-dimensional Gauss-Bonnet gravity.
However, in this case, to achieve stability, at least 3-dimensional
inner isotropic subspace is required \cite{Pavl1, Pavl2}. In the
six-dimensional case under consideration, we meet the stability
for the two-dimensional subspace. It is important that the
conformal terms help to stabilize extra dimensions even in the
linear gravity framework.

It is worth mentioning several possible ways to generalize the
present consideration. It is known that the Weyl anomaly consists
of three terms \cite{BFT-2000}, and the structure of the third
term is quite different from the two terms considered here. It
is possible to include the third term, but for this it is necessary
to adapt the scheme of isolating of highest derivative terms
since those are more complicated for the third term.

Furthermore,, the representation of five-dimensional spatial
metric in the form of a sum of the two isotropic subspaces
is postulated in this paper. However, it is known that in
Gauss-Bonnet gravity such a construction emerges naturally
from an initial totally anisotropic space (at least in the Bianchi I
geometry) during a cosmological evolution \cite{Chirkov}.
Whether the Weyl-enriched gravity may have the same
property, may be an interesting question for further research.

\section{Appendices}

\subsection{Weyl tensor scalars}
\label{weyl_scalars}

Let us give the expressions for the conformal terms in the
action for the metric which was used in the main part of our
manuscript. We have
{\small
\begin{align*}
& C_{\mu\nu \rho\si}C^{\rho\si \al\be}C_{\al\be}^{\,.\,.\,\mu\nu}
\, = \,
\frac{9}{50}\biggl\{ \frac{26e^{-6\gamma}}{3}-18\ddot{\gamma}
e^{-2\gamma-2\beta-2\alpha}+8\dot{\gamma}^{2}e^{-2\gamma
-2\beta-2\alpha}-34\dot{\beta}\dot{\gamma}e^{-2\gamma-2\beta-2\alpha}
\\
 &+4e^{-6\beta}
 +26\dot{\beta}^{2}e^{-2\gamma-2\beta-2\alpha}-18\dot{\alpha}
 \dot{\beta}e^{-2\gamma-2\beta-2\alpha}+4e^{-2\gamma-4\beta}
 +e^{-2\gamma-4\alpha}\biggl(4\ddot{\gamma}^{2}
 -18\dot{\gamma}^{2}\ddot{\gamma}
 \\
 & + 44\dot{\beta}\dot{\gamma}\ddot{\gamma}
 -44\dot{\beta}\ddot{\beta}\dot{\gamma}
 -8\dot{\alpha}\dot{\gamma}\ddot{\gamma}-8\ddot{\beta}\ddot{\gamma}
 -26\dot{\beta}^{2}\ddot{\gamma}+8\dot{\alpha}\dot{\beta}\ddot{\gamma}
 +4\dot{\gamma}^{4}-34\dot{\beta}\dot{\gamma}^{3}
 +18\dot{\alpha}\dot{\gamma}^{3}+18\ddot{\beta}\dot{\gamma}^{2}
 \\
 &+82\dot{\beta}^{2}\dot{\gamma}^{2}
 -62\dot{\alpha}\dot{\beta}\dot{\gamma}^{2}
 +4\dot{\alpha}^{2}\dot{\gamma}^{2} +8\dot{\alpha}\ddot{\beta}\dot{\gamma}
 -78\dot{\beta}^{3}\dot{\gamma}
 +70\dot{\alpha}\dot{\beta}^{2}\dot{\gamma}
 -8\dot{\alpha}^{2}\dot{\beta}\dot{\gamma}
 +4\ddot{\beta}^{2}+26\dot{\beta}^{2}\ddot{\beta}
 \\
&-8\dot{\alpha}\dot{\beta}\ddot{\beta}+26\dot{\beta}^{4}-26\dot{\alpha}\dot{\beta}^{3}
+4\dot{\alpha}^{2}\dot{\beta}^{2}\biggr) +e^{-4\gamma-2\alpha}\Bigl(-13\ddot{\gamma}+13\dot{\gamma}^{2}
-39\dot{\beta}\dot{\gamma}+13\dot{\alpha}\dot{\gamma}+13\ddot{\beta}
\\
&+26\dot{\beta}^{2}-13\dot{\alpha}\dot{\beta}\Bigr)
+e^{-6\alpha}\biggl(\frac{13\ddot{\gamma}^{3}}{3}
+17\dot{\gamma}^{2}\ddot{\gamma}^{2}
+76\dot{\alpha}\dot{\beta}\dot{\gamma}^{2}\ddot{\gamma}
-8\dot{\beta}^{2}\ddot{\beta}\ddot{\gamma}
-26\dot{\alpha}\dot{\beta}\ddot{\beta}\ddot{\gamma}
+34\dot{\alpha}\ddot{\beta}\dot{\gamma}^{3}
\\
 & -13\dot{\alpha}\dot{\gamma}\ddot{\gamma}^{2}
 -13\ddot{\beta}\ddot{\gamma}^{2}+4\dot{\beta}^{2}\ddot{\gamma}^{2}
 +13\dot{\alpha}\dot{\beta}\ddot{\gamma}^{2}
 +2\dot{\beta}\dot{\gamma}^{3}\ddot{\gamma}
 -34\dot{\alpha}\dot{\gamma}^{3}\ddot{\gamma}
 -34\ddot{\beta}\dot{\gamma}^{2}\ddot{\gamma}
 -41\dot{\beta}^{2}\dot{\gamma}^{2}\ddot{\gamma}
 \\
 & +13\dot{\alpha}^{2}\dot{\gamma}^{2}\ddot{\gamma}
 +42\dot{\beta}\ddot{\beta}\dot{\gamma}\ddot{\gamma}
 +26\dot{\alpha}\ddot{\beta}\dot{\gamma}\ddot{\gamma}
 +44\dot{\beta}^{3}\dot{\gamma}\ddot{\gamma}
 -50\dot{\alpha}\dot{\beta}^{2}\dot{\gamma}\ddot{\gamma}
 -26\dot{\alpha}^{2}\dot{\beta}\dot{\gamma}\ddot{\gamma}
 +13\ddot{\beta}^{2}\ddot{\gamma}
 \\
 & +8\dot{\alpha}\dot{\beta}^{3}\ddot{\gamma}
 +13\dot{\alpha}^{2}\dot{\beta}^{2}\ddot{\gamma}
 -16\dot{\beta}\dot{\gamma}^{5}-8\dot{\alpha}\dot{\gamma}^{5}
 -8\ddot{\beta}\dot{\gamma}^{4}
 +37\dot{\beta}^{2}\dot{\gamma}^{4}
 +6\dot{\alpha}\dot{\beta}\dot{\gamma}^{4}
 +17\dot{\alpha}^{2}\dot{\gamma}^{4}
 -2\dot{\beta}\ddot{\beta}\dot{\gamma}^{3}\\
 & -\frac{191\dot{\beta}^{3}\dot{\gamma}^{3}}{3}
 +43\dot{\alpha}\dot{\beta}^{2}\dot{\gamma}^{3}
 -55\dot{\alpha}^{2}\dot{\beta}\dot{\gamma}^{3}
 -\frac{13\dot{\alpha}^{3}\dot{\gamma}^{3}}{3}
 +17\ddot{\beta}^{2}\dot{\gamma}^{2}
 +41\dot{\beta}^{2}\ddot{\beta}\dot{\gamma}^{2}
 -76\dot{\alpha}\dot{\beta}\ddot{\beta}\dot{\gamma}^{2}
 +4\dot{\gamma}^{6}
 \\
 &-13\dot{\alpha}^{2}\ddot{\beta}\dot{\gamma}^{2}+26\dot{\alpha}^{2}\dot{\beta}\ddot{\beta}\dot{\gamma}-13\dot{\alpha}^{2}\dot{\beta}^{2}\ddot{\beta}+\frac{13\dot{\alpha}^{3}\dot{\beta}^{3}}{3}+\frac{26\dot{\beta}^{6}}{3}-39\dot{\beta}^{5}\dot{\gamma}+69\dot{\beta}^{4}\dot{\gamma}^{2}+8\dot{\gamma}^{4}\ddot{\gamma}\\
 & -85\dot{\alpha}\dot{\beta}^{3}\dot{\gamma}^{2}+63\dot{\alpha}^{2}\dot{\beta}^{2}\dot{\gamma}^{2}+13\dot{\alpha}^{3}\dot{\beta}\dot{\gamma}^{2}-21\dot{\beta}\ddot{\beta}^{2}\dot{\gamma}-13\dot{\alpha}\ddot{\beta}^{2}\dot{\gamma}-44\dot{\beta}^{3}\ddot{\beta}\dot{\gamma}+50\dot{\alpha}\dot{\beta}^{2}\ddot{\beta}\dot{\gamma}\\
 & +57\dot{\alpha}\dot{\beta}^{4}\dot{\gamma}-29\dot{\alpha}^{2}\dot{\beta}^{3}\dot{\gamma}-13\dot{\alpha}^{3}\dot{\beta}^{2}\dot{\gamma}-\frac{13\ddot{\beta}^{3}}{3}+4\dot{\beta}^{2}\ddot{\beta}^{2}+13\dot{\alpha}\dot{\beta}\ddot{\beta}^{2}+13\dot{\beta}^{4}\ddot{\beta}-8\dot{\alpha}\dot{\beta}^{3}\ddot{\beta}\\
 & -13\dot{\beta}^{4}\ddot{\gamma}-21\dot{\beta}\dot{\gamma}\ddot{\gamma}^{2}-13\dot{\alpha}\dot{\beta}^{5}
 + 4\dot{\alpha}^{2}\dot{\beta}^{4}\biggr)
 +e^{-2\beta-4\alpha}\Bigl(17\ddot{\gamma}^{2}+16\dot{\gamma}^{2}\ddot{\gamma}+2\dot{\beta}\dot{\gamma}\ddot{\gamma}-34\dot{\alpha}\dot{\gamma}\ddot{\gamma}\\
 & +12\dot{\gamma}^{4}-32\dot{\beta}\dot{\gamma}^{3}-16\dot{\alpha}\dot{\gamma}^{3}-16\ddot{\beta}\dot{\gamma}^{2}+41\dot{\beta}^{2}\dot{\gamma}^{2}+14\dot{\alpha}\dot{\beta}\dot{\gamma}^{2}+17\dot{\alpha}^{2}\dot{\gamma}^{2}-2\dot{\beta}\ddot{\beta}\dot{\gamma}+34\dot{\alpha}\ddot{\beta}\dot{\gamma}\\
 & -34\ddot{\beta}\ddot{\gamma}-34\dot{\beta}^{3}\dot{\gamma}+20\dot{\alpha}\dot{\beta}^{2}\dot{\gamma}-18\dot{\beta}^{2}\ddot{\gamma}+34\dot{\alpha}\dot{\beta}\ddot{\gamma}-34\dot{\alpha}^{2}\dot{\beta}\dot{\gamma}+18\dot{\beta}^{2}\ddot{\beta}-34\dot{\alpha}\dot{\beta}\ddot{\beta}+13\dot{\beta}^{4}\\
 & +17\dot{\alpha}^{2}\dot{\beta}^{2}\Bigl)+e^{-4\beta-2\alpha}
 \Bigl(8\ddot{\gamma}+12\dot{\gamma}^{2}-16\dot{\beta}\dot{\gamma}
 -8\dot{\alpha}\dot{\gamma} -8\ddot{\beta}+4\dot{\beta}^{2}+8\dot{\alpha}\dot{\beta}\Bigr)
 +17\ddot{\beta}^{2}-18\dot{\alpha}\dot{\beta}^{3}
 \\
 &+18\dot{\alpha}\dot{\gamma}e^{-2\gamma-2\beta-2\alpha}
 +18\ddot{\beta}e^{-2\gamma-2\beta-2\alpha}+13e^{-4\gamma-2\beta}
 \biggr\}
\end{align*}
and
\begin{align*}
 &
 C^{\mu \si \rho \nu}C_{\mu\al\be\nu}C^{\al\,\,\,\,\,\,\be}_{\,\,.\,\si\rho\,.}
 \,=\,
 \frac{9}{200}\biggl\{ -\frac{34e^{-6\gamma}}{3}-38\ddot{\gamma}e^{-2\gamma-2\beta-2\alpha}-72\dot{\gamma}^{2}e^{-2\gamma-2\beta-2\alpha}
 -36e^{-2\gamma-4\beta}\\
 &+106\dot{\beta}\dot{\gamma}e^{-2\gamma-2\beta-2\alpha} +38\ddot{\beta}e^{-2\gamma-2\beta-2\alpha}-34\dot{\beta}^{2}e^{-2\gamma-2\beta-2\alpha}-38\dot{\alpha}\dot{\beta}e^{-2\gamma-2\beta-2\alpha}+\bigg(4\dot{\beta}\dot{\gamma}\ddot{\gamma}\\
 & +72\dot{\alpha}\dot{\gamma}\ddot{\gamma}+72\ddot{\beta}\ddot{\gamma}+34\dot{\beta}^{2}\ddot{\gamma}-72\dot{\alpha}\dot{\beta}\ddot{\gamma}+106\dot{\beta}\dot{\gamma}^{3}+38\dot{\alpha}\dot{\gamma}^{3}+38\ddot{\beta}\dot{\gamma}^{2}-138\dot{\beta}^{2}\dot{\gamma}^{2}-42\dot{\alpha}\dot{\beta}\dot{\gamma}^{2}\\
 & -4\dot{\beta}\ddot{\beta}\dot{\gamma}-72\dot{\alpha}\ddot{\beta}\dot{\gamma}+102\dot{\beta}^{3}\dot{\gamma}-30\dot{\alpha}\dot{\beta}^{2}\dot{\gamma}+72\dot{\alpha}^{2}\dot{\beta}\dot{\gamma}-36\ddot{\beta}^{2}-34\dot{\beta}^{2}\ddot{\beta}+72\dot{\alpha}\dot{\beta}\ddot{\beta}+34\dot{\alpha}\dot{\beta}^{3}\\
 &
-34\dot{\beta}^{4}-36\dot{\gamma}^{4}-36\ddot{\gamma}^{2}-38\dot{\gamma}^{2}\ddot{\gamma}-36\dot{\alpha}^{2}\dot{\beta}^{2}-36\dot{\alpha}^{2}\dot{\gamma}^{2}\bigg)e^{-2\gamma-4\alpha}+e^{-4\gamma-2\alpha}\left(-17\dot{\gamma}^{2}+51\dot{\beta}\dot{\gamma}\right.\\
&\left.+17\ddot{\gamma}-17\dot{\alpha}\dot{\gamma}-17\ddot{\beta}-34\dot{\beta}^{2}+17\dot{\alpha}\dot{\beta}\right)+e^{-6\alpha}\left(-\frac{17\ddot{\gamma}^{3}}{3}-53\dot{\gamma}^{2}\ddot{\gamma}^{2}-131\dot{\beta}^{2}\dot{\gamma}^{2}\ddot{\gamma}+72\dot{\beta}^{2}\ddot{\beta}\ddot{\gamma}\right.\\
 & +89\dot{\beta}\dot{\gamma}\ddot{\gamma}^{2}+17\dot{\alpha}\dot{\gamma}\ddot{\gamma}^{2}+17\ddot{\beta}\ddot{\gamma}^{2}-36\dot{\beta}^{2}\ddot{\gamma}^{2}-17\dot{\alpha}\dot{\beta}\ddot{\gamma}^{2}+182\dot{\beta}\dot{\gamma}^{3}\ddot{\gamma}+106\dot{\alpha}\dot{\gamma}^{3}\ddot{\gamma}+131\dot{\beta}^{2}\ddot{\beta}\dot{\gamma}^{2}\\
 & -284\dot{\alpha}\dot{\beta}\dot{\gamma}^{2}\ddot{\gamma}-17\dot{\alpha}^{2}\dot{\gamma}^{2}\ddot{\gamma}-178\dot{\beta}\ddot{\beta}\dot{\gamma}\ddot{\gamma}-34\dot{\alpha}\ddot{\beta}\dot{\gamma}\ddot{\gamma}+4\dot{\beta}^{3}\dot{\gamma}\ddot{\gamma}+250\dot{\alpha}\dot{\beta}^{2}\dot{\gamma}\ddot{\gamma}+34\dot{\alpha}^{2}\dot{\beta}\dot{\gamma}\ddot{\gamma}\\
 & +34\dot{\alpha}\dot{\beta}\ddot{\beta}\ddot{\gamma}+17\dot{\beta}^{4}\ddot{\gamma}-72\dot{\alpha}\dot{\beta}^{3}\ddot{\gamma}-17\dot{\alpha}^{2}\dot{\beta}^{2}\ddot{\gamma}-36\dot{\gamma}^{6}+144\dot{\beta}\dot{\gamma}^{5}+72\dot{\alpha}\dot{\gamma}^{5}+72\ddot{\beta}\dot{\gamma}^{4}-53\ddot{\beta}^{2}\dot{\gamma}^{2}\\
 & -182\dot{\beta}\ddot{\beta}\dot{\gamma}^{3}-106\dot{\alpha}\ddot{\beta}\dot{\gamma}^{3}+\frac{619\dot{\beta}^{3}\dot{\gamma}^{3}}{3}+313\dot{\alpha}\dot{\beta}^{2}\dot{\gamma}^{3}+195\dot{\alpha}^{2}\dot{\beta}\dot{\gamma}^{3}+\frac{17\dot{\alpha}^{3}\dot{\gamma}^{3}}{3}-233\dot{\beta}^{2}\dot{\gamma}^{4}\\
 & +17\dot{\alpha}^{2}\ddot{\beta}\dot{\gamma}^{2}-121\dot{\beta}^{4}\dot{\gamma}^{2}-135\dot{\alpha}\dot{\beta}^{3}\dot{\gamma}^{2}-17\dot{\alpha}^{3}\dot{\beta}\dot{\gamma}^{2}+89\dot{\beta}\ddot{\beta}^{2}\dot{\gamma}+17\dot{\alpha}\ddot{\beta}^{2}\dot{\gamma}-4\dot{\beta}^{3}\ddot{\beta}\dot{\gamma}-17\dot{\alpha}\dot{\beta}\ddot{\beta}^{2}\\
 & -267\dot{\alpha}^{2}\dot{\beta}^{2}\dot{\gamma}^{2}-34\dot{\alpha}^{2}\dot{\beta}\ddot{\beta}\dot{\gamma}+51\dot{\beta}^{5}\dot{\gamma}-13\dot{\alpha}\dot{\beta}^{4}\dot{\gamma}+161\dot{\alpha}^{2}\dot{\beta}^{3}\dot{\gamma}+17\dot{\alpha}^{3}\dot{\beta}^{2}\dot{\gamma}+\frac{17\ddot{\beta}^{3}}{3}-36\dot{\beta}^{2}\ddot{\beta}^{2}\\
 & +72\dot{\alpha}\dot{\beta}^{3}\ddot{\beta}-250\dot{\alpha}\dot{\beta}^{2}\ddot{\beta}\dot{\gamma}+284\dot{\alpha}\dot{\beta}\ddot{\beta}\dot{\gamma}^{2}+17\dot{\alpha}^{2}\dot{\beta}^{2}\ddot{\beta}-\frac{34\dot{\beta}^{6}}{3}-36\dot{\alpha}^{2}\dot{\beta}^{4}-\frac{17\dot{\alpha}^{3}\dot{\beta}^{3}}{3}-17\ddot{\beta}^{2}\ddot{\gamma}\\
 &
-254\dot{\alpha}\dot{\beta}\dot{\gamma}^{4}-72\dot{\gamma}^{4}\ddot{\gamma}-17\dot{\beta}^{4}\ddot{\beta}+17\dot{\alpha}\dot{\beta}^{5}-53\dot{\alpha}^{2}\dot{\gamma}^{4}+106\ddot{\beta}\dot{\gamma}^{2}\ddot{\gamma}\bigg)+e^{-2\beta-4\alpha}\Big(-53\dot{\alpha}^{2}\dot{\gamma}^{2}\\
 & -38\dot{\beta}^{2}\ddot{\gamma}-106\dot{\alpha}\dot{\beta}\ddot{\gamma}-108\dot{\gamma}^{4}+288\dot{\beta}\dot{\gamma}^{3}+144\ddot{\beta}\dot{\gamma}^{2}-269\dot{\beta}^{2}\dot{\gamma}^{2}-326\dot{\alpha}\dot{\beta}\dot{\gamma}^{2}-182\dot{\beta}\ddot{\beta}\dot{\gamma}\\
 & -106\dot{\alpha}\ddot{\beta}\dot{\gamma}+106\dot{\beta}^{3}\dot{\gamma}+220\dot{\alpha}\dot{\beta}^{2}\dot{\gamma}+106\dot{\alpha}^{2}\dot{\beta}\dot{\gamma}+38\dot{\beta}^{2}\ddot{\beta}+106\dot{\alpha}\dot{\beta}\ddot{\beta}-38\dot{\alpha}\dot{\beta}^{3}-53\dot{\alpha}^{2}\dot{\beta}^{2}\\
 &-53\ddot{\beta}^{2}-53\ddot{\gamma}^{2}+144\dot{\alpha}\dot{\gamma}^{3}-144\dot{\gamma}^{2}\ddot{\gamma}-17\dot{\beta}^{4}182\dot{\beta}\dot{\gamma}\ddot{\gamma}+106\dot{\alpha}\dot{\gamma}\ddot{\gamma}+106\ddot{\beta}\ddot{\gamma}\Big)-17e^{-4\gamma-2\beta}\\
 & +e^{-4\beta-2\alpha}\Big(-72\ddot{\gamma}-108\dot{\gamma}^{2} +144\dot{\beta}\dot{\gamma}+72\dot{\alpha}\dot{\gamma}+72\ddot{\beta}-36\dot{\beta}^{2}-72\dot{\alpha}\dot{\beta}\Big)-36e^{-6\beta}\\
 &+38\dot{\alpha}\dot{\gamma}e^{-2\gamma-2\beta-2\alpha}\biggr\}
\end{align*}
}

\newpage

\subsection{Field Equations }

Here we collect the formulas related to variational derivatives.

\subsubsection{Variation in $\alpha$ \label{F_alpha}}

{\small
\begin{align*}
& F_{\alpha}=\frac{9}{50}\biggl\{ -\frac{25\Lambda}{36}+ \theta_{1}e^{-2\gamma}\left(-36\dot{\beta}\dot{\gamma}\ddot{\gamma}
-18\ddot{\beta}\ddot{\gamma}+36\dot{\beta}^{2}\ddot{\gamma}
+27\dot{\gamma}^{4}-108\dot{\beta}\dot{\gamma}^{3}
-18\ddot{\beta}\dot{\gamma}^{2}+9\ddot{\gamma}^{2}\right.\\
& \left.
-54\dot{\beta}^{2}\ddot{\beta}+72\dot{\beta}\ddot{\beta}\dot{\gamma}
-54\dot{\beta}^{3}\dot{\gamma}+9\ddot{\beta}^{2}
+135\dot{\beta}^{2}\dot{\gamma}^{2}\right)
+ \theta_{2}e^{-2\gamma}\left(-4\ddot{\beta}^{2}
+16\dot{\beta}\dot{\gamma}\ddot{\gamma}-16\dot{\beta}^{2}\ddot{\gamma}
-12\dot{\gamma}^{4}\right.\\
 &
 \left.+8\ddot{\beta}\ddot{\gamma}+8\ddot{\beta}\dot{\gamma}^{2}
 -4\ddot{\gamma}^{2}-60\dot{\beta}^{2}\dot{\gamma}^{2}
 -32\dot{\beta}\ddot{\beta}\dot{\gamma}
 +24\dot{\beta}^{3}\dot{\gamma}
 +24\dot{\beta}^{2}\ddot{\beta}
 +48\dot{\beta}\dot{\gamma}^{3}\right)
 -\frac{17 \theta_{1}e^{-4\gamma-2\beta}}{4}\\
 &
 -\frac{17 \theta_{1}e^{-6\gamma}}{6}+18 \theta_{1}e^{-2\gamma-2\beta}\left(\dot{\gamma}-\dot{\beta}\right)^{2}
 -8 \theta_{2}e^{-2\gamma-2\beta}\left(\dot{\gamma}-\dot{\beta}\right)^{2}
 +4 \theta_{2}e^{-2\gamma-4\beta}-9 \theta_{1}e^{-2\gamma-4\beta}\\
 &
 + \theta_{1}
 \bigg(
 -\frac{87\dot{\gamma}^{2}\ddot{\gamma}^{2}}{4}
 +\frac{53\dot{\beta}\dot{\gamma}\ddot{\gamma}^{2}}{2}
 -\frac{17\ddot{\beta}\ddot{\gamma}^{2}}{2}
 -\frac{19\dot{\beta}^{2}\ddot{\gamma}^{2}}{4}
 -53\dot{\gamma}^{4}\ddot{\gamma}
 +89\dot{\beta}\dot{\gamma}^{3}\ddot{\gamma}
+35\ddot{\beta}\dot{\gamma}^{2}\ddot{\gamma}
-36\dot{\beta}\ddot{\beta}\dot{\gamma}\ddot{\gamma}
\\
 & +\frac{17\ddot{\beta}^{2}\ddot{\gamma}}{2}
 +\dot{\beta}^{2}\ddot{\beta}\ddot{\gamma}
 +36\dot{\beta}^{4}\ddot{\gamma}-107\dot{\beta}\dot{\gamma}^{5}
 +\frac{53\ddot{\beta}\dot{\gamma}^{4}}{2}
 +\frac{1301\dot{\beta}^{2}\dot{\gamma}^{4}}{4}
 +\frac{17\dot{\beta}\ddot{\beta}\dot{\gamma}^{3}}{2}
 -\frac{2551\dot{\beta}^{3}\dot{\gamma}^{3}}{6}\\
 &
+17\dot{\beta}^{2}\dot{\gamma}^{2}\ddot{\gamma}
-89\dot{\beta}^{3}\dot{\gamma}\ddot{\gamma}
-\frac{53\ddot{\beta}^{2}\dot{\gamma}^{2}}{4}
-\frac{301\dot{\beta}^{2}\ddot{\beta}\dot{\gamma}^{2}}{2}
+259\dot{\beta}^{4}\dot{\gamma}^{2}
+\frac{19\dot{\beta}\ddot{\beta}^{2}\dot{\gamma}}{2}
+\frac{339\dot{\beta}^{3}\ddot{\beta}\dot{\gamma}}{2}
+9\dot{\gamma}^{6}\\
 &
 -\frac{125\dot{\beta}^{5}\dot{\gamma}}{2}
 +\frac{17\ddot{\gamma}^{3}}{6}-\frac{17\ddot{\beta}^{3}}{6}
 +\frac{15\dot{\beta}^{2}\ddot{\beta}^{2}}{4}
 -54\dot{\beta}^{4}\ddot{\beta}+\frac{17\dot{\beta}^{6}}{12}\bigg)
 + \theta_{2}\biggl(16\dot{\beta}\ddot{\beta}\dot{\gamma}\ddot{\gamma} -35\dot{\beta}^{2}\ddot{\beta}^{2}-\frac{13\dot{\beta}^{6}}{3}
 \nonumber
 \\
 & -\frac{26\ddot{\gamma}^{3}}{3}+43\dot{\gamma}^{2}\ddot{\gamma}^{2}
 -34\dot{\beta}\dot{\gamma}\ddot{\gamma}^{2}+26\ddot{\beta}\ddot{\gamma}^{2}+68\dot{\gamma}^{4}\ddot{\gamma}-84\dot{\beta}\dot{\gamma}^{3}\ddot{\gamma}-60\ddot{\beta}\dot{\gamma}^{2}\ddot{\gamma}-52\dot{\beta}^{2}\dot{\gamma}^{2}\ddot{\gamma}-289\dot{\beta}^{2}\dot{\gamma}^{4}\nonumber\\
 & -9\dot{\beta}^{2}\ddot{\gamma}^{2}+17\ddot{\beta}^{2}\dot{\gamma}^{2}+84\dot{\beta}^{3}\dot{\gamma}\ddot{\gamma}-26\ddot{\beta}^{2}\ddot{\gamma}-16\dot{\beta}^{4}\ddot{\gamma}+92\dot{\beta}\dot{\gamma}^{5}-34\ddot{\beta}\dot{\gamma}^{4}+24\dot{\beta}^{4}\ddot{\beta}-4\dot{\gamma}^{6}+\frac{26\ddot{\beta}^{3}}{3}\nonumber\\
 &
 +44\dot{\beta}^{2}\ddot{\beta}\ddot{\gamma}
 -26\dot{\beta}\ddot{\beta}\dot{\gamma}^{3}
 +\frac{1078\dot{\beta}^{3}\dot{\gamma}^{3}}{3}
 -204\dot{\beta}^{4}\dot{\gamma}^{2}
 +178\dot{\beta}^{2}\ddot{\beta}\dot{\gamma}^{2}
 +18\dot{\beta}\ddot{\beta}^{2}\dot{\gamma}
 -142\dot{\beta}^{3}\ddot{\beta}\dot{\gamma}
 +50\dot{\beta}^{5}\dot{\gamma}\bigg)
 \\
 &+ \theta_{1}\biggl(\frac{53\ddot{\gamma}^{2}}{4} -53\dot{\gamma}^{2}\ddot{\gamma}+53\dot{\beta}\dot{\gamma}\ddot{\gamma}-\frac{53\ddot{\beta}\ddot{\gamma}}{2}-89\dot{\beta}\dot{\gamma}^{3}+\frac{53\ddot{\beta}\dot{\gamma}^{2}}{2}+\frac{693\dot{\beta}^{2}\dot{\gamma}^{2}}{4}-\frac{231\dot{\beta}^{3}\dot{\gamma}}{2}+\frac{53\ddot{\beta}^{2}}{4}\nonumber\\
 & +9\dot{\gamma}^{4}-\frac{53\dot{\beta}^{2}\ddot{\beta}}{2}+\frac{89\dot{\beta}^{4}}{4}\biggr)e^{-2\beta} +\theta_{2}e^{-2\beta}\Bigl(68\dot{\gamma}^{2}\ddot{\gamma}-17\ddot{\gamma}^{2}-68\dot{\beta}\dot{\gamma}\ddot{\gamma}+34\ddot{\beta}\ddot{\gamma}+84\dot{\beta}\dot{\gamma}^{3}-4\dot{\gamma}^{4}\nonumber\\
&-34\ddot{\beta}\dot{\gamma}^{2}-177\dot{\beta}^{2}\dot{\gamma}^{2}+118\dot{\beta}^{3}\dot{\gamma}-17\ddot{\beta}^{2}+34\dot{\beta}^{2}\ddot{\beta}-21\dot{\beta}^{4}\Bigr) -4 \theta_{2}e^{-4\beta}\left(\dot{\gamma}-\dot{\beta}\right)^{2}+\frac{25e^{-2\beta}}{12\pi G}\nonumber \\
 & -9 \theta_{1}e^{-6\beta}+\frac{25\dot{\beta}\dot{\gamma}}{6\pi G}-\frac{17 \theta_{1}}{4}e^{-4\gamma}\left(\dot{\gamma}-\dot{\beta}\right)^{2}+13 \theta_{2}e^{-4\gamma-2\beta}
 +13 \theta_{2}e^{-4\gamma}\left(\dot{\gamma}-\dot{\beta}\right)^{2}+4 \theta_{2}e^{-6\beta}\nonumber\\
 &
 +\frac{25e^{-2\gamma}}{36\pi G}+\frac{26 \theta_{2}e^{-6\gamma}}{3}+\frac{25\dot{\beta}^{2}}{12\pi G}+\frac{25\dot{\gamma}^{2}}{36\pi G}-9 \theta_{1}e^{-4\beta}\left(\dot{\gamma}-\dot{\beta}\right)^{2}\biggr\} \nonumber
\end{align*}
}

\subsubsection{Variation in $\beta$ \label{F_beta}}

{\small
\begin{align*}
 & F_{\beta}=\frac{9}{50}\left\{ -\frac{25\Lambda}{12\pi G}{}+\frac{25\ddot{\beta}}{6\pi G}+ \theta_{1}e^{-2\gamma}\Bigl(18\dot{\gamma}\dddot{\gamma}+90\dot{\beta}\dddot{\gamma}-108\dot{\gamma}^{2}\ddot{\gamma}+252\dot{\beta}\dot{\gamma}\ddot{\gamma}+90\ddot{\beta}\ddot{\gamma}+54\dot{\beta}^{2}\ddot{\gamma}\right.\\
 & -27\dot{\gamma}^{4}+126\ddot{\beta}\dot{\gamma}^{2}+189\dot{\beta}^{2}\dot{\gamma}^{2}-162\dot{\beta}\ddot{\beta}\dot{\gamma}-162\dot{\beta}^{3}\dot{\gamma}-108\dot{\beta}\dddot{\beta}-81\ddot{\beta}^{2}-162\dot{\beta}^{2}\ddot{\beta}-9\ddot{\gamma}^{2}\Bigr)\\
 &+ \theta_{2}e^{-2\gamma}\left(-40\dot{\beta}\dddot{\gamma}+4\ddot{\gamma}^{2}-112\dot{\beta}\dot{\gamma}\ddot{\gamma}+48\dot{\gamma}^{2}\ddot{\gamma}-40\ddot{\beta}\ddot{\gamma}-24\dot{\beta}^{2}\ddot{\gamma}+12\dot{\gamma}^{4}-56\ddot{\beta}\dot{\gamma}^{2}-84\dot{\beta}^{2}\dot{\gamma}^{2}\right.\\
 &
+72\dot{\beta}\ddot{\beta}\dot{\gamma}+72\dot{\beta}^{3}\dot{\gamma}+48\dot{\beta}\dddot{\beta} +36\ddot{\beta}^{2}+72\dot{\beta}^{2}\ddot{\beta}-8\dot{\gamma}\dddot{\gamma}\Bigr)+8 \theta_{2}\left(2\ddot{\gamma}+\dot{\gamma}^{2}-2\ddot{\beta}-\dot{\beta}^{2}\right)e^{-2\gamma-2\beta}\\
&+\frac{25e^{-2\gamma}}{12\pi G}+13 \theta_{2}\Bigl(\dot{\gamma}^{2}-2\ddot{\gamma}-4\dot{\beta}\dot{\gamma}+2\ddot{\beta}+3\dot{\beta}^{2}\Bigr)e^{-4\gamma}
 -\frac{17 \theta_{1}e^{-6\gamma}}{2}-\frac{17 \theta_{1}e^{-4\gamma-2\beta}}{4} \nonumber\\
 &+\frac{17}{4} \theta_{1}\left(34\ddot{\gamma}-17\dot{\gamma}^{2}+68\dot{\beta}\dot{\gamma}-34\ddot{\beta}-51\dot{\beta}^{2}\right)e^{-4\gamma}+18 \theta_{1}e^{-2\gamma-2\beta}\Bigl(\dot{\beta}^{2}-2\ddot{\gamma}-\dot{\gamma}^{2}+2\ddot{\beta}\Bigl)\\
 &-4 \theta_{2}e^{-2\gamma-4\beta}+26 \theta_{2}e^{-6\gamma}+13 \theta_{2}e^{-4\gamma-2\beta}+9 \theta_{1}e^{-2\gamma-4\beta} + \theta_{1}\Biggl(\frac{297\dot{\gamma}\ddot{\gamma}\dddot{\gamma}}{2}-\frac{713\dot{\beta}\ddot{\beta}\dot{\gamma}\ddot{\gamma}}{2}\\
 &-\frac{93\dot{\beta}\ddot{\gamma}\dddot{\gamma}}{2}+\frac{265\dot{\gamma}^{3}\dddot{\gamma}}{2}-90\dot{\beta}\dot{\gamma}^{2}\dddot{\gamma}-\frac{265\dot{\beta}^{2}\dot{\gamma}\dddot{\gamma}}{2}+\frac{1109\dot{\gamma}^{2}\ddot{\gamma}^{2}}{4}-27\dot{\gamma}^{6}+37\ddot{\beta}^{2}\ddot{\gamma}+19\dot{\beta}\dddot{\beta}\dot{\gamma}^{2}\\
 & +\frac{17\dddot{\gamma}^{2}}{2}-17\dddot{\beta}\dddot{\gamma}+\frac{59\dot{\beta}\ddot{\beta}\dddot{\gamma}}{2}+90\dot{\beta}^{3}\dddot{\gamma}+35\ddot{\gamma}^{3}-4\dot{\beta}\dot{\gamma}\ddot{\gamma}^{2}-71\ddot{\beta}\ddot{\gamma}^{2}-\frac{413\dot{\beta}^{2}\ddot{\gamma}^{2}}{4}+105\dot{\gamma}^{4}\ddot{\gamma}\\
 & +532\dot{\beta}\dot{\gamma}^{3}\ddot{\gamma}-300\ddot{\beta}\dot{\gamma}^{2}\ddot{\gamma}-1087\dot{\beta}^{2}\dot{\gamma}^{2}\ddot{\gamma}-140\dddot{\beta}\dot{\gamma}\ddot{\gamma}+\frac{809\dot{\beta}^{3}\dot{\gamma}\ddot{\gamma}}{2}+38\dot{\beta}\dddot{\beta}\ddot{\gamma}+\frac{633\dot{\beta}^{2}\ddot{\beta}\ddot{\gamma}}{2}-\ddot{\beta}^{3}\\
 & +\frac{91\dot{\beta}^{4}\ddot{\gamma}}{2}+267\dot{\beta}\dot{\gamma}^{5}+\frac{55\ddot{\beta}\dot{\gamma}^{4}}{2}-\frac{1551\dot{\beta}^{2}\dot{\gamma}^{4}}{4}-106\dddot{\beta}\dot{\gamma}^{3}-728\dot{\beta}\ddot{\beta}\dot{\gamma}^{3}-88\dot{\beta}^{3}\dot{\gamma}^{3}+\frac{125\ddot{\beta}^{2}\dot{\gamma}^{2}}{4}\\
 & +\frac{1947\dot{\beta}^{2}\ddot{\beta}\dot{\gamma}^{2}}{2}+\frac{821\dot{\beta}^{4}\dot{\gamma}^{2}}{2}+123\ddot{\beta}\dddot{\beta}\dot{\gamma}+195\dot{\beta}^{2}\dddot{\beta}\dot{\gamma}+\frac{687\dot{\beta}\ddot{\beta}^{2}\dot{\gamma}}{2}-\frac{239\dot{\beta}^{3}\ddot{\beta}\dot{\gamma}}{2}-179\dot{\beta}^{5}\dot{\gamma}\\
 &+\frac{17\dddot{\beta}^{2}}{2}-108\dot{\beta}^{3}\dddot{\beta}-21\dot{\beta}\ddot{\beta}\dddot{\beta}\Biggr)
 +4 \theta_{1}\left(-819\dot{\beta}^{2}\ddot{\beta}^{2}-614\dot{\beta}^{4}\ddot{\beta}+17\dot{\beta}^{6}\right)+ \theta_{2}\Bigl(52\dddot{\beta}\dddot{\gamma}\\
&+40\dot{\beta}\dot{\gamma}^{2}\dddot{\gamma}+214\ddot{\beta}\dot{\gamma}\dddot{\gamma}-60\ddot{\gamma}^{3}-592\dot{\beta}\dot{\gamma}^{3}\ddot{\gamma}+57\dot{\beta}^{2}\ddot{\gamma}^{2}-252\dot{\beta}\dot{\gamma}^{5}-188\ddot{\beta}\dddot{\beta}\dot{\gamma}-338\dot{\beta}^{4}\dot{\gamma}^{2}\\
 & -170\dot{\gamma}^{3}\dddot{\gamma}+170\dot{\beta}^{2}\dot{\gamma}\dddot{\gamma}+98\dot{\beta}\ddot{\beta}\dddot{\gamma}-401\dot{\gamma}^{2}\ddot{\gamma}^{2}-176\dot{\beta}\dot{\gamma}\ddot{\gamma}^{2}-202\dot{\beta}^{3}\dot{\gamma}\ddot{\gamma}+400\ddot{\beta}\dot{\gamma}^{2}\ddot{\gamma}+12\dot{\gamma}^{6}\\
 & +339\dot{\beta}^{2}\dot{\gamma}^{4}+76\ddot{\beta}\ddot{\gamma}^{2}+972\dot{\beta}^{2}\dot{\gamma}^{2}\ddot{\gamma}+240\dddot{\beta}\dot{\gamma}\ddot{\gamma}+714\dot{\beta}\ddot{\beta}\dot{\gamma}\ddot{\gamma}+72\dot{\beta}\dddot{\beta}\ddot{\gamma}+28\ddot{\beta}^{2}\ddot{\gamma}-74\dot{\beta}^{2}\ddot{\beta}\ddot{\gamma}\\
 & -58\dot{\beta}^{3}\ddot{\beta}\dot{\gamma}-40\dot{\beta}^{3}\dddot{\gamma}+136\dddot{\beta}\dot{\gamma}^{3}+768\dot{\beta}\ddot{\beta}\dot{\gamma}^{3}+128\dot{\beta}^{3}\dot{\gamma}^{3}+36\dot{\beta}\dddot{\beta}\dot{\gamma}^{2}-25\ddot{\beta}^{2}\dot{\gamma}^{2}-766\dot{\beta}^{2}\ddot{\beta}\dot{\gamma}^{2}\\
 & -220\dot{\beta}^{2}\dddot{\beta}\dot{\gamma}+10\ddot{\beta}\dot{\gamma}^{4}-486\dot{\beta}\ddot{\beta}^{2}\dot{\gamma}+124\dot{\beta}^{5}\dot{\gamma}-26\dddot{\beta}^{2}-124\dot{\beta}\ddot{\beta}\dddot{\beta}+48\dot{\beta}^{3}\dddot{\beta}-44\ddot{\beta}^{3}-9\dot{\beta}^{2}\ddot{\beta}^{2}\\
 &-46\dot{\beta}\ddot{\gamma}\dddot{\gamma}+46\dot{\beta}^{4}\ddot{\beta}-180\dot{\gamma}^{4}\ddot{\gamma}-266\dot{\gamma}\ddot{\gamma}\dddot{\gamma}+2\dot{\beta}^{4}\ddot{\gamma}-26\dddot{\gamma}^{2}-13\dot{\beta}^{6}\Bigl) + \theta_{1}\Biggl(\frac{265\dot{\gamma}\dddot{\gamma}}{2}+\frac{53\dot{\beta}\dddot{\gamma}}{2}\\
 &+284\dot{\beta}\dot{\gamma}\ddot{\gamma}-\frac{53\ddot{\beta}\ddot{\gamma}}{2}-\frac{231\dot{\beta}^{2}\ddot{\gamma}}{2}-27\dot{\gamma}^{4} +231\dot{\beta}\dot{\gamma}^{3}+\frac{19\ddot{\beta}\dot{\gamma}^{2}}{2}-\frac{799\dot{\beta}^{2}\dot{\gamma}^{2}}{4}-106\dddot{\beta}\dot{\gamma}-\frac{727\dot{\beta}\ddot{\beta}\dot{\gamma}}{2}\\
 &+\frac{265\ddot{\gamma}^{2}}{4}
+123\dot{\gamma}^{2}\ddot{\gamma}-\frac{53\dot{\beta}^{3}\dot{\gamma}}{2}-53\dot{\beta}\dddot{\beta}-\frac{159\ddot{\beta}^{2}}{4}+\frac{125\dot{\beta}^{2}\ddot{\beta}}{2}+\frac{89\dot{\beta}^{4}}{4}\Biggr)e^{-2\beta}
 + \theta_{2}e^{-2\beta}\Bigl( -85\ddot{\gamma}^{2}\\
&
+34\ddot{\beta}\ddot{\gamma}+118\dot{\beta}^{2}\ddot{\gamma}-236\dot{\beta}\dot{\gamma}^{3}+18\ddot{\beta}\dot{\gamma}^{2}+211\dot{\beta}^{2}\dot{\gamma}^{2}+406\dot{\beta}\ddot{\beta}\dot{\gamma}+136\dddot{\beta}\dot{\gamma}-170\dot{\gamma}\dddot{\gamma}-34\dot{\beta}\dddot{\gamma}
 \end{align*}
\begin{align*}
&-188\dot{\gamma}^{2}\ddot{\gamma}-304\dot{\beta}\dot{\gamma}\ddot{\gamma}+12\dot{\gamma}^{4}+34\dot{\beta}^{3}\dot{\gamma}+68\dot{\beta}\dddot{\beta}+51\ddot{\beta}^{2}-50\dot{\beta}^{2}\ddot{\beta}-21\dot{\beta}^{4}\Bigl)+9 \theta_{1}e^{-4\beta}\Bigr(+3\dot{\gamma}^{2}\\
&-4\dot{\beta}\dot{\gamma}+2\ddot{\gamma}-2\ddot{\beta}+9\dot{\beta}\Bigr)
+\frac{25\dot{\gamma}^{2}}{4\pi G}+\frac{25\dot{\beta}\dot{\gamma}}{3\pi G}+4 \theta_{2}e^{-4\beta}\left(-2\ddot{\gamma}-3\dot{\gamma}^{2}+4\dot{\beta}\dot{\gamma}+2\ddot{\beta}-\dot{\beta}^{2}\right)\nonumber\\
 &-12 \theta_{2}e^{-6\beta}+27 \theta_{1}e^{-6\beta}+\frac{25\dot{\beta}^{2}}{4\pi G}+\frac{25e^{-2\beta}}{12\pi G}+\frac{25\ddot{\gamma}}{6\pi G}\Biggr\} \nonumber
\end{align*}
}

\subsubsection{Variation in $\gamma$}
\label{F_gamma}

{\small
\begin{align*}
&F_\gamma=\frac{9}{50}\left\{- \frac{25\Lambda}{18}+ \theta_{2}e^{-2\gamma}\left(48\dot{\beta}\dddot{\gamma}-48\dot{\gamma}^{2}\ddot{\gamma}+96\dot{\beta}\dot{\gamma}\ddot{\gamma}+32\ddot{\beta}\ddot{\gamma}+40\dot{\beta}^{2}\ddot{\gamma}-32\ddot{\beta}^{2}+8\dddot{\beta}\dot{\gamma}\right.\right.\\
 & \left.-48\dot{\beta}\dot{\gamma}^{3}+48\ddot{\beta}\dot{\gamma}^{2}+144\dot{\beta}^{2}\dot{\gamma}^{2}-40\dot{\beta}\ddot{\beta}\dot{\gamma}-96\dot{\beta}^{3}\dot{\gamma}-56\dot{\beta}\dddot{\beta}-96\dot{\beta}^{2}\ddot{\beta}\right)+ \theta_{1}e^{-2\gamma}\Bigl(-72\ddot{\beta}\ddot{\gamma}\\
 & +108\dot{\gamma}^{2}\ddot{\gamma}-216\dot{\beta}\dot{\gamma}\ddot{\gamma}-108\dot{\beta}\dddot{\gamma}-90\dot{\beta}^{2}\ddot{\gamma}+108\dot{\beta}\dot{\gamma}^{3}-108\ddot{\beta}\dot{\gamma}^{2}-324\dot{\beta}^{2}\dot{\gamma}^{2}-18\dddot{\beta}\dot{\gamma}+90\dot{\beta}\ddot{\beta}\dot{\gamma}\\
 & \left.+216\dot{\beta}^{3}\dot{\gamma}+126\dot{\beta}\dddot{\beta}+216\dot{\beta}^{2}\ddot{\beta}+72\ddot{\beta}^{2}\right)+ \theta_{2}e^{-4\gamma}\left(26\ddot{\gamma}-26\dot{\gamma}^{2}+78\dot{\beta}\dot{\gamma}-26\ddot{\beta}-52\dot{\beta}^{2}\right)\\
 &
 \frac{17 \theta_{1}}{2}\Bigl(-\ddot{\gamma}+\dot{\gamma}^{2}-3\dot{\beta}\dot{\gamma}+\ddot{\beta}+2\dot{\beta}^{2}\Bigr)e^{-4\gamma}-\frac{104 \theta_{2}e^{-6\gamma}}{3}+\frac{17 \theta_{1}e^{-4\gamma-2\beta}}{2}-26 \theta_{2}e^{-4\gamma-2\beta}\\
 & +36 \theta_{1}\left(\ddot{\gamma}+\dot{\beta}\dot{\gamma}-\ddot{\beta}-\dot{\beta}^{2}\right)e^{-2\gamma-2\beta}+16 \theta_{2}\left(-\ddot{\gamma}-\dot{\beta}\dot{\gamma}+\ddot{\beta}+\dot{\beta}^{2}\right)e^{-2\gamma-2\beta}+ \theta_{2}\Biggl(26\dddot{\gamma}^{2}\\
&+72\dot{\beta}\ddot{\gamma}\dddot{\gamma}+136\dot{\gamma}^{3}\dddot{\gamma}+36\dot{\beta}\dot{\gamma}^{2}\dddot{\gamma}-188\ddot{\beta}\dot{\gamma}\dddot{\gamma}-52\dddot{\beta}\dddot{\gamma}-124\dot{\beta}\ddot{\beta}\dddot{\gamma}+48\dot{\beta}^{3}\dddot{\gamma}-920\dot{\beta}^{2}\dot{\gamma}^{2}\ddot{\gamma}\\
 & +\frac{206\ddot{\gamma}^{3}}{3}+358\dot{\gamma}^{2}\ddot{\gamma}^{2}+210\dot{\beta}\dot{\gamma}\ddot{\gamma}^{2}-102\ddot{\beta}\ddot{\gamma}^{2}-48\dot{\beta}^{2}\ddot{\gamma}^{2}+112\dot{\gamma}^{4}\ddot{\gamma}+676\dot{\beta}\dot{\gamma}^{3}\ddot{\gamma}-220\dot{\beta}^{2}\dot{\gamma}\dddot{\gamma}\\
 & -214\dddot{\beta}\dot{\gamma}\ddot{\gamma}-730\dot{\beta}\ddot{\beta}\dot{\gamma}\ddot{\gamma}+118\dot{\beta}^{3}\dot{\gamma}\ddot{\gamma}-98\dot{\beta}\dddot{\beta}\ddot{\gamma}-2\ddot{\beta}^{2}\ddot{\gamma}+30\dot{\beta}^{2}\ddot{\beta}\ddot{\gamma}+14\dot{\beta}^{4}\ddot{\gamma}+160\dot{\beta}\dot{\gamma}^{5}\\
 & -742\dot{\beta}\ddot{\beta}\dot{\gamma}^{3}-\frac{1462\dot{\beta}^{3}\dot{\gamma}^{3}}{3}-112\dot{\beta}\dddot{\beta}\dot{\gamma}^{2}+588\dot{\beta}^{2}\ddot{\beta}\dot{\gamma}^{2}+542\dot{\beta}^{4}\dot{\gamma}^{2}+162\ddot{\beta}\dddot{\beta}\dot{\gamma}+\frac{106\ddot{\beta}^{3}}{3}\\
 &
+8\ddot{\beta}^{2}\dot{\gamma}^{2}-174\dot{\beta}^{5}\dot{\gamma}-102\dddot{\beta}\dot{\gamma}^{3}+468\dot{\beta}\ddot{\beta}^{2}\dot{\gamma}+200\dot{\beta}^{3}\ddot{\beta}\dot{\gamma}+26\dddot{\beta}^{2}+150\dot{\beta}\ddot{\beta}\dddot{\beta}-56\dot{\beta}^{3}\dddot{\beta}\\
&-8\dot{\gamma}^{6}+44\dot{\beta}^{2}\ddot{\beta}^{2}-70\dot{\beta}^{4}\ddot{\beta}+\frac{52\dot{\beta}^{6}}{3}+240\dot{\gamma}\ddot{\gamma}\dddot{\gamma}-50\dot{\beta}^{2}\dot{\gamma}^{4}-340\ddot{\beta}\dot{\gamma}^{2}\ddot{\gamma}+24\ddot{\beta}\dot{\gamma}^{4}+270\dot{\beta}^{2}\dddot{\beta}\dot{\gamma}\Biggr)\\
&+ \theta_{1}\Biggl(+1070\dot{\beta}^{2}\dot{\gamma}^{2}\ddot{\gamma}-140\dot{\gamma}\ddot{\gamma}\dddot{\gamma}+\frac{415\dot{\beta}^{4}\ddot{\beta}}{2}-\frac{17\dot{\beta}^{6}}{3}+38\dot{\beta}\ddot{\gamma}\dddot{\gamma}+\frac{785\dot{\beta}\ddot{\beta}\dot{\gamma}\ddot{\gamma}}{2}+\frac{125\dot{\beta}^{2}\dot{\gamma}^{4}}{2}\\
 & -106\dot{\gamma}^{3}\dddot{\gamma}+19\dot{\beta}\dot{\gamma}^{2}\dddot{\gamma}+195\dot{\beta}^{2}\dot{\gamma}\dddot{\gamma}+-21\dot{\beta}\ddot{\beta}\dddot{\gamma}-108\dot{\beta}^{3}\dddot{\gamma}-\frac{227\ddot{\gamma}^{3}}{6}-\frac{511\dot{\gamma}^{2}\ddot{\gamma}^{2}}{2}-\frac{45\dot{\beta}\dot{\gamma}\ddot{\gamma}^{2}}{2}\\
 & +123\ddot{\beta}\dot{\gamma}\dddot{\gamma}+\frac{159\ddot{\beta}\ddot{\gamma}^{2}}{2}+108\dot{\beta}^{2}\ddot{\gamma}^{2}-52\dot{\gamma}^{4}\ddot{\gamma}-621\dot{\beta}\dot{\gamma}^{3}\ddot{\gamma}+265\ddot{\beta}\dot{\gamma}^{2}\ddot{\gamma}+\frac{263\dddot{\beta}\dot{\gamma}\ddot{\gamma}}{2}-\frac{631\dot{\beta}^{3}\dot{\gamma}\ddot{\gamma}}{2}\\
 & -\frac{17\dddot{\gamma}^{2}}{2}-\frac{59\dot{\beta}\dddot{\beta}\ddot{\gamma}}{2}-\frac{91\ddot{\beta}^{2}\ddot{\gamma}}{2}-\frac{635\dot{\beta}^{2}\ddot{\beta}\ddot{\gamma}}{2}-\frac{163\dot{\beta}^{4}\ddot{\gamma}}{2}+18\dot{\gamma}^{6}-160\dot{\beta}\dot{\gamma}^{5}-54\ddot{\beta}\dot{\gamma}^{4}+17\dddot{\beta}\dddot{\gamma}\\
 & +\frac{1439\dot{\beta}\ddot{\beta}\dot{\gamma}^{3}}{2}+\frac{3079\dot{\beta}^{3}\dot{\gamma}^{3}}{6}-18\ddot{\beta}^{2}\dot{\gamma}^{2}-823\dot{\beta}^{2}\ddot{\beta}\dot{\gamma}^{2}-\frac{1339\dot{\beta}^{4}\dot{\gamma}^{2}}{2}-\frac{229\ddot{\beta}\dddot{\beta}\dot{\gamma}}{2}-\frac{515\dot{\beta}^{2}\dddot{\beta}\dot{\gamma}}{2}\\
 & -353\dot{\beta}\ddot{\beta}^{2}\dot{\gamma}+\frac{159\dddot{\beta}\dot{\gamma}^{3}}{2}+52\dot{\beta}\dddot{\beta}\dot{\gamma}^{2}+\frac{483\dot{\beta}^{5}\dot{\gamma}}{2}-\frac{17\dddot{\beta}^{2}}{2}+\frac{25\dot{\beta}\ddot{\beta}\dddot{\beta}}{2}+126\dot{\beta}^{3}\dddot{\beta}+\frac{23\ddot{\beta}^{3}}{6}
 \end{align*}
\begin{align*}
 &+201\dot{\beta}^{2}\ddot{\beta}^{2}-50\dot{\beta}^{3}\ddot{\beta}\dot{\gamma}\Biggr)+\frac{25\dot{\beta}\dot{\gamma}}{6\pi G}+\frac{25\dot{\beta}^{2}}{3\pi G}+\frac{34 \theta_{1}e^{-6\gamma}}{3}+8 \theta_{2}e^{-6\beta}+\frac{25\dot{\gamma}^{2}}{18\pi G}\\
 &+ \theta_{2}\Bigl(136\dot{\gamma}\dddot{\gamma}-406\dot{\beta}\ddot{\beta}\dot{\gamma}+68\dot{\beta}\dddot{\gamma}-152\dot{\beta}^{3}\dot{\gamma}-102\dot{\beta}\dddot{\beta}-34\ddot{\beta}^{2}+16\dot{\beta}^{2}\ddot{\beta}+372\dot{\beta}\dot{\gamma}\ddot{\gamma}-8\dot{\gamma}^{4}\\
 & +102\ddot{\gamma}^{2}+120\dot{\gamma}^{2}\ddot{\gamma}-68\ddot{\beta}\ddot{\gamma}-118\dot{\beta}^{2}\ddot{\gamma}+152\dot{\beta}\dot{\gamma}^{3}+16\ddot{\beta}\dot{\gamma}^{2}-34\dot{\beta}^{2}\dot{\gamma}^{2}-102\dddot{\beta}\dot{\gamma}+42\dot{\beta}^{4}\Bigr)e^{-2\beta}\\
 &+\frac{25\ddot{\gamma}}{18\pi G}+ \theta_{1}e^{-2\beta}\left(-106\dot{\gamma}\dddot{\gamma}-53\dot{\beta}\dddot{\gamma}-70\dot{\gamma}^{2}\ddot{\gamma}-337\dot{\beta}\dot{\gamma}\ddot{\gamma}+53\ddot{\beta}\ddot{\gamma}+\frac{231\dot{\beta}^{2}\ddot{\gamma}}{2}-\frac{159\ddot{\gamma}^{2}}{2}\right.\\
 & -142\dot{\beta}\dot{\gamma}^{3}-36\ddot{\beta}\dot{\gamma}^{2}+\frac{53\dot{\beta}^{2}\dot{\gamma}^{2}}{2}+\frac{159\dddot{\beta}\dot{\gamma}}{2}+\frac{727\dot{\beta}\ddot{\beta}\dot{\gamma}}{2}+142\dot{\beta}^{3}\dot{\gamma}+\frac{159\dot{\beta}\dddot{\beta}}{2}+\frac{53\ddot{\beta}^{2}}{2}-36\dot{\beta}^{2}\ddot{\beta}\nonumber\\
 & -\frac{89\dot{\beta}^{4}}{2}+18\dot{\gamma}^{4}\Biggr)+\frac{25\ddot{\beta}}{6\pi G}+8 \theta_{2}e^{-4\beta}\left(\ddot{\gamma}+\dot{\gamma}^{2}-\dot{\beta}\dot{\gamma}-\ddot{\beta}\right)+18 \theta_{1}e^{-4\beta}\left(-\ddot{\gamma}-\dot{\gamma}^{2}+\dot{\beta}\dot{\gamma}+\ddot{\beta}\right)\nonumber\\
 & +\frac{25e^{-2\beta}}{6\pi G}-18 \theta_{1}e^{-6\beta}\Biggr\} \nonumber
\end{align*}
}

\section*{Acknowledgements}
I.Sh. is grateful to CNPq (Conselho Nacional de Desenvolvimento
Cient\'{i}fico e Tecnol\'{o}gico, Brazil)  for the partial support
under grant 305122/2023-1.
The work of S.R. is funded by the Ministry of Science and Higher
Education of the Russian Federation, Project "New Phenomena in
Particle Physics and the Early Universe" FSWU-2023-0073.


\end{document}